\documentclass[prd,showpacs,nofootinbib,preprint, 10 pt]{revtex4}
\usepackage{amssymb}
\usepackage{amsmath,graphicx,color,epsfig}
\usepackage{pstricks}
\usepackage{float}

\newcommand{\en}{\enspace}

\newcommand{\BBfitmassY}{10.90}
\newcommand{\BBfitmassYMeV}{10900}
\newcommand{\BBfitwidthY}{28 }
\newcommand{\BBConstmassY}{10.890 }
\newcommand{\BBConstmassYTwo}{11.130}
\newcommand{\kappaval}{0.87}
\newcommand{\kappavalError}{0.13}
\newcommand{\chsqfit}{88/67 }
\newcommand{\tn}[1]{\textnormal{#1}}

\newcommand{\wQ}{\cal Q}
\begin{document}

\begin{flushright}
DESY 09-192\\
November 2009
\end{flushright}

\title{A Case for Hidden  $\mathbf{b \bar{b}}$ Tetraquarks based on
$\mathbf{ e^+e^- \to b \bar{b}}$ Cross Section between $\sqrt{s}=10.54$
and $\mathbf 11.20$ GeV}

\author{Ahmed~Ali}
\email{ahmed.ali@desy.de}
\affiliation{Deutsches Elektronen-Synchrotron DESY, D-22607 Hamburg, Germany}

\author{Christian Hambrock}
\email{christian.hambrock@desy.de}
\affiliation{Deutsches Elektronen-Synchrotron DESY, D-22607 Hamburg, Germany}

\author{Ishtiaq Ahmed}
\email{ishtiaq.ahmed@ncp.edu.pk}
\affiliation{National Centre for Physics, Quaid-i-Azam, University, Islamabad,
 Pakistan}

\author{M. Jamil Aslam}
\email{muhammadjamil.aslam@gmail.com}
\affiliation{Physics Department, Quaid-i-Azam, University, Islamabad,
 Pakistan}

\date{\today}

\begin{abstract}
We study the spectroscopy and dominant decays of the bottomonium-like 
tetraquarks (bound diquarks-antidiquarks), focusing on the lowest lying
P-wave  $[bq][\bar{b}\bar{q}]$ states $Y_{[bq]}$ 
 (with $q=u,d$), having $J^{PC}=1^{--}$. To search for them, we analyse the
BaBar data~\cite{:2008hx} obtained during an energy scan of the $e ^+ e^- \to b \bar{b}$
 cross section in the range of $\sqrt{s}=10.54$ to 11.20 GeV. 
 We find that these data are consistent with the presence of an
additional $b \bar{b}$ state $Y_{[bq]}$ with a mass of 10.90 GeV
and a width of about 30 MeV  apart from the $\Upsilon (5S)$ and $\Upsilon(6S)$
 resonances. A closeup of the energy region around the
 $Y_{[bq]}$-mass may resolve this state in terms of the two mass eigenstates,
 $Y_{[b,l]}$ and $Y_{[b,h]}$, with a mass difference, estimated as about 6 MeV.
 We tentatively identify the state $Y_{[bq]}(10900)$
from the $R_b$-scan with the state $Y_b(10890)$ observed by
 Belle~\cite{Abe:2007tk}  in the process
 $e^+e^- \to Y_b(10890) \to \Upsilon (1S, 2S)\;\pi^+ \pi^-$ due to their
 proximity in  masses and decay widths. 

\end{abstract}

\maketitle

\section{Introduction}
In the past several years, experiments at the two B-factories, BaBar and Belle, and at the Tevatron
collider, CDF and D0, have discovered an impressive number of new hadronic states
in the mass region of the charmonia~\cite{Amsler:2008zzb}. 
These states generically labelled as $X$, $Y$ and $Z$, however, defy a conventional
$c\bar{c}$ charmonium interpretation~ \cite{Jaffe:2004ph,Quigg:2004vf}. Moreover, they
 are quite numerous, with some 14 of them discovered by the last count, ranging in mass
from the $J^{PC}=1^{++}$ $X(3872)$, decaying into $D \bar{D}^*, J/\psi \pi^+\pi^-, J\psi \gamma$, to
the $J^{PC}=1^{--}$ $Y(4660)$, decaying into $\psi^\prime \pi^+\pi^-$ (for a recent 
experimental summary and references, see ~\cite{Zupanc:2009qc}).
 There is also evidence for an $s\bar{s}$ bound state, $Y_s(2175)$ having the
quantum numbers $J^{PC}=1^{--}$, first observed by BaBar in the initial state radiation (ISR) process
$e^+ e^- \to \gamma_{\rm ISR}\;f_0(980)\phi(1020)$, where $f_0(980)$ is the
$0^{++}$ scalar state~\cite{Aubert:2006bu}. This was later confirmed by BES~\cite{:2007yt}
and Belle~\cite{Shen:2009zze}.

These states are the subject of intense phenomenological
 studies. Three different frameworks have been suggested to accommodate them: (i) 
$D-D^{\ast }$ molecules \cite{Tornqvist:2004qy,Liu:2005ay,Rosner:2007mu};
 (ii) $c\bar{c}g$ hybrids \cite{Kou:2005gt}; and (iii) Diquark-antidiquark or
four quark states~\cite{Maiani:2004vq,Maiani:2005pe,Drenska:2008gr}. Of these hypotheses (i) and
(iii) are more popular. For example, 
the motivation to explain the state $X\left( 3872\right)$, first observed by Belle~\cite{Choi:2003ue}and later confirmed by 
CDF~\cite{Acosta:2003zx}, D0~\cite{Abazov:2004kp} and BaBar~\cite{Aubert:2004ns},
as a hadronic molecule
is that the mass of this state is very close to the $D^{0}\bar{D}^{\ast 0}$ threshold.
 Hence, in this picture,
the binding energy is small implying that these are not compact hadrons, which have 
typical sizes of $O(1)$ Fermi.
This makes it unlikely that such a loosely bound state could be produced promptly 
(i.e. not from $B$ decays,
 as seen by Belle and BaBar) in high energy hadron collisions, unless one tailors the
wave functions to avoid this conclusion. In particular,
 Bignamini et al.~\cite{Bignamini:2009sk} have estimated the prompt production
cross section of $X\left( 3872\right) $ at the Tevatron, assuming it as a $D^{0}\bar{D}%
^{\ast 0}$ hadron molecule.
Their upper bound on the cross section $p\bar{p} \to X(3872)+...$  is about two orders of
magnitude smaller than the minimum production cross section
from the CDF data~\cite{Abulencia:2006ma}, disfavouring the molecular interpretation of $X(3872)$.
However, a dissenting estimate~\cite{Artoisenet:2009} yields a much larger
cross section, invoking the charm meson rescatterings.

The case  that the $X, Y, Z$ and $Y_s$ are diquark-antidiquark  hadrons,
in which the diquark (antidiquark) pairs are in colour $\bar{3}_c$ ($3_c$) configuration
bound together by the QCD colour forces, has been forcefully
made by Maiani, Polosa and their collaborators~\cite{Maiani:2004vq,Maiani:2005pe,Drenska:2008gr}. 
 The idea itself that diquarks in this colour configuration 
can play a fundamental role in hadron spectroscopy is rather old, going back well over thirty years 
to the suggestions by Jaffe~\cite{Jaffe:1976ih}. More recently, diquarks were revived  by
Jaffe and Wilczek~\cite{Jaffe:2003sg} in the context of exotic hadron spectroscopy, in particular,
pentaquark baryons (antidiquark-antidiquark-quark), which now seem to have receded into oblivion.
However, diquarks as constituents of hadronic matter may (eventually) find their
rightful place in particle physics.   Lately, interest in
this proposal has re-emerged, with a well-founded theoretical interpretation of the low lying scalar mesons
as dominantly diquark-antidiquark states and the ones lying higher in mass in the
1 - 2 GeV region as being dominantly $q\bar{q}$ mesons~\cite{Hooft:2008we}.  Evidence in favour
of an attractive diquark (antidiquark) $qq$ channel for the so-called {\it good} diquarks
 (colour antitriplet
$\bar{3}_c$, flavour antisymmetric $ \bar{3}_f$, spin-singlet positive parity) in the characterisation
of Jaffe~\cite{Jaffe:2004ph} is now also emerging from more than one Lattice QCD
 studies~\cite{Alford:2000mm,Alexandrou:2006cq} for the light quark systems.
 On the other hand, no evidence is found on the lattice for an attractive diquark channel for the
so-called {\it bad} diquarks (i.e., spin-1 states) involving light quarks~\cite{Alexandrou:2006cq}. However, as the
effective QCD Lagrangian is  spin-independent in the heavy quark limit, we anticipate that
 also the {\it bad} diquarks will be found to be in attractive channel
 for the $[cq]$ and $[bq]$
 diquarks having a charm or a beauty quark. This implies a huge number
of heavy tetraquark states, as we also show here
for the hidden $b\bar{b}$ tetraquark spectroscopy.  Earlier work along these lines has
 been reported in the literature using relativistic quark models~\cite{Ebert:2008se} and
 QCD sum rules~\cite{Wang:2009kw}. 

In this paper, we study the tetraquark picture in the bottom
($b\bar{b}$)  sector. In the first part (Section II), we classify these states 
according to their $J^{PC}$ quantum numbers and calculate the mass spectrum of the
diquarks-antidiquarks  $[bq][\bar{b} \bar{q}^{\prime }]$ with $q$, $q^{\prime }=u$, $d$,
 $s$ and $c$  in the ground and
orbitally excited states by assuming both {\it good} and {\it bad} diquarks. The resulting
mass spectrum for the $0^{++}, 1^{++}, 1^{+-}, 1^{--}$ and $2^{++}$ states  having the valence 
diquark-antidiquark content  $[bq][\bar{b}\bar{q}]$, with $q=u,d, s$ and $c$, and the mixed ones
$[bd][\bar{b} \bar{s}]$ (and charge conjugates) is shown in Fig.~\ref{FullSpectrum}.
The main focus of this letter is on the $J^{PC}=1^{--}$ states, which are excited
$P$-wave states. To be specific, there are four neutral states 
$Y_{[bu]}^{(n)}$ $(n=1,...,4)$  with the quark content $([bu][\bar{b}\bar{u}])$ (which differ in their spin
 assignments) and
another four $Y_{[bd]}^{(n)}$ with the quark content
 $([bd][\bar{b}\bar{d}])$. In the isospin symmetry limit, which is
used in calculating the entries in Fig.~\ref{FullSpectrum}, these mass states are degenerate for
each $n$. Isospin-breaking introduces a mass
 splitting and the mass eigenstates called $Y^{(n)}_{[b, l]}$ and $Y^{(n)}_{[b, h]}$ (for 
lighter and heavier of the two) become linear combinations of $Y_{[bu]}^{(n)}$ and
$Y_{[bd]}^{(n)}$. Thus,  $Y^{(n)}_{[b, l]}\equiv \cos \theta\; Y^{(n)}_{[bu]} +
 \sin \theta\; Y^{(n)}_{[bd]}$
 and $Y^{(n)}_{[b, h]}\equiv -\sin \theta \; Y^{(n)}_{[bu]} + \cos \theta \; Y^{(n)}_{[bd]}$.
The mass differences are estimated to be small, with 
 $M(Y^{(n)}_{[b, h]}) -M(Y_{[b, l]})= (7 \pm 2)  \cos 2\theta$ MeV, where $\theta$
 is a mixing angle. The electromagnetic couplings of the tetraquarks
$Y^{(n)}_{[b,l]}$ and $Y^{(n)}_{[b,h]}$ are calculated assuming
that the diquarks have point-like couplings with the photon, given  by
 $eQ_{[bq]}$ where $e^2/(4\pi)$ is the electromagnetic
fine structure constant $\alpha$ and $Q_{[bq]}= +1/3$  for
the $[bu]$ and $[bc]$ diquarks and $Q_{[bq]}=-2/3$ for the $[bd]$ and $[bs]$ diquarks.
Because of this charge assignment,  electromagnetic couplings of the tetraquarks 
$Y^{(n)}_{[b,l]}$ and $Y^{(n)}_{[b,h]}$ will depend on the mixing angle $\theta$ (Section III).

To calculate the production cross sections $e^+ e^- \to Y^{(n)}_{[b, l]} \to {\rm hadrons}$ and
$e^+ e^- \to Y^{(n)}_{[b, h]} \to {\rm hadrons}$,
 we need to calculate the partial widths $\Gamma^{(n)}_{ee}(Y_{[b,l]})$
and  $\Gamma^{(n)}_{ee}(Y_{[b,h]})$ for decays  into $e^{+}e^{-}$ pair and
 the hadronic decay widths  $\Gamma(Y^{(n)}_{[b,l]})$ and $\Gamma(Y^{(n)}_{[b,h]})$.
 For the $\Upsilon (nS)$, 
 the leptonic decay widths are determined by the wave functions at the origin
$\Psi_{b\bar{b}}(0)$. The tetraquark states  $Y^{(n)}_{[b,l]}$ and $Y^{(n)}_{[b,h]}$
are P-wave states, and we need  the derivative of the corresponding wave functions 
 at the origin, $\Psi^\prime_{b\bar{b}}(0)$. 
 To take into account the possibly larger hadronic
 size of the tetraquarks compared to  that of the $b\bar{b}$ mesons, we modify the Quarkonia
 potential, usually taken as a sum of linear (confining) and Coulombic (short-distance) parts.
 For example, the Buchm\"uller-Tye $Q\bar{Q}$ potential~\cite{Buchmuller:1980su} has the
asymptotic forms $V(r) \sim k_{Q\bar{Q}} \;r$ (for $r \to \infty $) and
 $V(r) \sim 1/r\ln(1/\Lambda_{\rm QCD}^2\;r^2)$ (for $r \to 0 $),
where $k_{Q\bar{Q}}$ is the string tension and $\Lambda_{\rm QCD}$ is the QCD
scale  parameter. The bound state tetraquark potential
$V_{\wQ \bar{\wQ}}(r)$\footnote{We shall use the symbol $\wQ$ and $\bar{\wQ}$ to denote a 
generic diquark and antidiquark, respectively. However, where the flavour content of the diquark
 is to be specified, we use the symbol $[bq]$, and $[\bar{b}\bar{q}]$ with $q=u,d,s,c$.}
 will differ from the Quarkonia potential
 $V_{Q\bar{Q}}(r)$ in the linear part, as the string tension in a diquark $k_{\wQ \wQ}$ is
expected to be different than the corresponding string tension $k_{Q\bar{Q}}$
 in the $Q\bar{Q}$ mesons, but
as the diquarks-antidiquarks in the tetraquarks and the quarks-antiquarks in the mesons are
 in the same ($\bar{3}_c 3_c$) colour
 configuration, the Coulomb (short-distance) parts of the potentials will be similar. 
 Defining $\kappa=k_{\wQ\bar{\wQ}}/k_{Q\bar{Q}}$, we expect $\kappa$ to have a value
in the range  $\kappa\in [\frac{1}{2},\frac{\sqrt{3}}{2}]$~\cite{Alexandrou:2006cq}.
 A value of $\kappa$ different from unity will modify the tetraquark wave functions
 $\Psi_{\wQ\bar{\wQ}}(0)$ from the corresponding ones of the bound $b\bar{b}$ systems,
 effecting the leptonic decay widths of the tetraquarks. 
Hadronic decays of $Y^{(n)}_{[b,l]}$ and $Y^{(n)}_{[b,h]}$ are calculated by relating them to
 the corresponding decays of the $\Upsilon(5S)$, such as
$\Upsilon(5S) \to B^{(*)} \bar{B}^{(*)}$, which we take from the PDG. 
 We assume that the form factors in the two set of decays $(Y_{[b,q]}$ and $\Upsilon (5S)$) are related
by $\kappa$, yielding the hadronic decay widths (Section IV).

Having specified the mass spectrum and our dynamical assumptions for the tetraquark decays, we undertake
  a theoretical analysis of the existing data
from BaBar~\cite{:2008hx} on 
$R_b(s)= \sigma(e^+ e^- \to b\bar{b})/\sigma(e^+ e^- \to \mu^+ \mu^-)$,
obtained during an energy scan of the $e ^+ e^- \to b \bar{b}$ cross section
in the range of $\sqrt{s}=10.54$ to 11.20 GeV. The question that we ask and try to
partially answer is:
Are the kinematically allowed tetraquark states $Y^{(n)}_{[b, h]}$ and $Y^{(n)}_{[b, l]}$ visible
 in the BaBar energy scan of $R_b$?
 To that end, we calculate the contributions of the lowest $1^{--}$  tetraquark states
$Y_{[b, h]}$ and $Y_{[b, l]}$ to the hadronic cross sections $\sigma(e^+ e^- \to Y_{[b,l]} \to \;{\rm hadrons})$ 
and $\sigma(e^+ e^- \to Y_{[b,h]} \to \:{\rm hadrons})$, and hence the corresponding contributions 
 $\Delta R_b(s)$.\footnote{We shall often refer to the ground states $Y^{(1)}_{[b, h]}$ and
 $Y^{(1)}_{[b, l]}$ without the superscript for ease of writing.}
  Our fits of the BaBar
$R_b$-data are consistent with the presence of a single state $Y_{[bq]}$
as a Breit-Wigner resonance with the mass
around $\BBfitmassY$~GeV and a width of about $30$ MeV, in addition to the $\Upsilon(5S)$ and
 $\Upsilon(6S)$.   The quality of the fit with three Breit-Wigners is found to be  better 
 than the one obtained with just 2 (i.e., with $\Upsilon(5S)$ and $\Upsilon(6S)$),
 as reported by BaBar~\cite{:2008hx} (Section V). 
A closeup of the energy region around \BBfitmassY~GeV is necessary to confirm and
resolve the structure reported by us, as the isospin-induced mass difference between the two
 eigenstates $Y_{[b, h]}$ and $Y_{[b, l]}$ comes out as about 6 MeV, which 
 is comparable to the  BaBar centre-of-mass energy step of 5 MeV. We hope that this can be
investigated in the near future by Belle. 

 We tentatively identify the state $Y_{[bq]}(10900)$ with the state $Y_b(10890)$ measured in the
process $e^+ e^- \to Y_b(10890) \to \Upsilon(1S,2S)\; \pi^+\pi^-$~\cite{Abe:2007tk}. 
 An analysis~\cite{Ali:2009es} of the Belle data on the decay widths
 $\Gamma(Y_b \to  \Upsilon(1S,2S)\; \pi^+\pi^-)$, dipion invariant mass spectra and
the helicity angular distributions is in agreement with the tetraquark interpretation presented here.
%

\section{Spectrum of bottom diquark-antidiquark states}
The mass spectrum of tetraquarks $[bq][\overline{bq^{\prime }]}$ with $q=u$, 
$d$, $s$ and $c$ can be described in terms of the constituent diquark
masses, $m_{\wQ}$, spin-spin interactions inside the single diquark, spin-spin
interaction between quark and antiquark belonging to two diquarks,
spin-orbit, and purely orbital term \cite{Drenska:2008gr}, i.e.
\begin{equation}
H=2m_{\wQ}+H_{SS}^{(\wQ\wQ)}+H_{SS}^{(\wQ\bar{\wQ}\mathcal{)}}+H_{SL}+H_{LL} ,
\label{01}
\end{equation}%
where:%
\begin{eqnarray}
H_{SS}^{(\wQ\wQ)} &=&2(\mathcal{K}_{bq})_{\bar{3}}[(\mathbf{S}_{b}\cdot \mathbf{S%
}_{q})+(\mathbf{S}_{\bar{b}}\cdot \mathbf{S}_{\bar{q}})],  \notag \\
H_{SS}^{(\wQ\bar{\wQ}\mathcal{)}} &=&2(\mathcal{K}_{b\bar{q}})(\mathbf{S}%
_{b}\cdot \mathbf{S}_{\bar{q}}+\mathbf{S}_{\bar{b}}\cdot \mathbf{S}_{q})+2%
\mathcal{K}_{b\bar{b}}(\mathbf{S}_{b}\cdot \mathbf{S}_{\bar{b}})+2\mathcal{K}%
_{q\bar{q}}(\mathbf{S}_{q}\cdot \mathbf{S}_{\bar{q}}),  \notag \\
H_{SL} &=&2A_{\wQ}(\mathbf{S}_{\mathcal{Q}}\cdot \mathbf{L}+\mathbf{S}_{%
\mathcal{\bar{Q}}}\cdot \mathbf{L}),  \notag \\
H_{LL} &=&B_{\wQ}\frac{L_{\wQ\bar{\wQ}}(L_{\wQ\bar{\wQ}}+1)}{2}.  \label{02}
\end{eqnarray}%
Here $m_{\wQ}$ is the mass of the diquark $[bq]$, $(\mathcal{K}_{bq})_{\bar{3}%
} $ is the spin-spin interaction between the quarks inside the diquarks, $%
\mathcal{K}_{b\bar{q}}$ are the couplings ranging outside the diquark
shells, $A_{\wQ}$ is the spin-orbit coupling of diquark and $B_{\wQ}$
corresponds to the contribution of the total angular momentum of the
diquark-antidiquark system to its mass. The overall factor of $2$ is used
customarily in the literature. For the calculation of the masses we assume isospin
symmetry, i.e. the isodoublet consisting of the states
\begin{equation}
\label{def_high_low_states}
Y^{(n)}_{[bu]}=[ b u ] [ \bar{b} \bar{u} ]\en\en\en\textnormal{and} \en\en\en
  Y^{(n)}_{[bd]}=[ b d ] [ \bar{b} \bar{d} ]
\end{equation}
are degenerate in mass for each $n$. Later,  we
will calculate the isospin symmetry breaking effects in the masses. 

The parameters involved in the above Hamiltonian (\ref{02}) can be obtained
from the known meson and baryon masses by resorting to the constituent quark
model~\cite{De Rujula:1975ge}
\begin{equation}
H=\sum\limits_{i}m_{i}+\sum\limits_{i<j}2\mathcal{K}_{ij}(\mathbf{S}%
_{i}\cdot \mathbf{S}_{j}) , \label{Constituent-model}
\end{equation}%
where the sum runs over the hadron constituents. The coefficient $\mathcal{K}%
_{ij}$ depends on the flavour of the constituents $i$, $j$ and on the particular
colour state of the pair. Using the entries in the PDG for hadron masses along
with the assumption that the spin-spin interactions are independent of
whether the quarks belong to a meson or a diquark, the results for diquark
masses corresponding to $X\left( 3872\right) $ and $Y\left( 2175\right) $
were calculated in the literature \cite{Maiani:2004vq,Drenska:2008gr}. Here, we
extend this  procedure to the tetraquarks $[bq][\bar{b}\bar{q}]$.
The constituent quark masses and the couplings $%
\mathcal{K}_{ij}$\ for the colour singlet and antitriplet states are given in
Table I, II and III. 
\begin{table}[tb]
\caption{Constituent quark masses derived from the $L=0$ mesons and baryons.}
\label{Table I}
\begin{center}
\begin{tabular}{|l|l|l|l|l|}
\hline
Constituent mass (MeV) & $q$ & $s$ & $c$ & $b$ \\ \hline
Mesons & $305$ & $490$ & $1670$ & $5008$ \\ \hline
Baryons & $362$ & $546$ & $1721$ & $5050$ \\ \hline\hline
\end{tabular}%
\end{center}
\end{table}

\begin{table}[tb]
\caption{Spin-Spin couplings for quark-antiquark pairs in the colour
singlet state from the known mesons.}
\label{Table II}
\begin{center}
\begin{tabular}{|l|l|l|l|l|l|l|l|l|l|l|}
\hline
Spin-spin couplings & $q\bar{q}$ & $s\bar{q}$ & $s\bar{s}$ & $c\bar{q}$ & $c%
\bar{s}$ & $c\bar{c}$ & $b\bar{q}$ & $b\bar{s}$ & $b\bar{c}$ & $b\bar{b}$ \\ 
\hline
$\left( \mathcal{K}_{ij}\right) _{0}$(MeV) & $318$ & $200$ & $129$ & $71$ & $%
72$ & $59$ & $23$ & $23$ & $20$ & $36$ \\ \hline\hline
\end{tabular}%
\end{center}
\end{table}
\begin{table}[tb]
\caption{Spin-Spin couplings for quark-quark pairs in colour $\bar{3}$ state from the 
known baryons.}
\label{Table III}
\begin{center}
\begin{tabular}{|l|l|l|l|l|l|l|l|l|}
\hline
Spin-Spin couplings & $qq$ & $sq$ & $cq$ & $cs$ & $ss$ & $bq$ & $bs$ & $bc$
\\ \hline
$\left( \mathcal{K}_{ij}\right) _{\bar{3}}$(MeV) & $98$ & $65$ & $22$ & $24$
& $72$ & $6$ & $25$ & $10$ \\ \hline\hline
\end{tabular}%
\end{center}
\end{table}
To calculate the spin-spin interaction of the $\wQ\bar{\wQ}$ states
explicitly, we use the non-relativistic notation 
$\left\vert S_{\wQ}\text{,~}S_{\bar{\wQ}};~J\right\rangle $,
where $S_{\wQ}$ and $S_{\bar{\wQ}}$ are the spin of diquark and antidiquark, respectively,
and $J$ is the total angular momentum. These states are then defined in terms of the
direct product of the $2\times 2$ matrices in spinor space, $\Gamma ^{\alpha }$, which 
can be written in terms of the Pauli matrices as: 
\begin{equation}
\Gamma ^{0}=\frac{\sigma _{2}}{\sqrt{2}};~\Gamma ^{i}=\frac{1}{\sqrt{2}}%
\sigma _{2}\sigma _{i}~  ,\label{05}
\end{equation}%
which then lead to the following definitions: 
\begin{eqnarray}
\left\vert 0_{\wQ},0_{\bar{\wQ}};~0_{J}\right\rangle &=&\frac{1}{2}\left( \sigma
_{2}\right) \otimes \left( \sigma _{2}\right) ,  \notag \\
\left\vert 1_{\wQ},1_{\bar{\wQ}};~0_{J}\right\rangle &=&\frac{1}{2\sqrt{3}}%
\left( \sigma _{2}\sigma ^{i}\right) \otimes \left( \sigma _{2}\sigma
^{i}\right) ,  \notag \\
\left\vert 0_{\wQ},1_{\bar{\wQ}};~1_{J}\right\rangle &=&\frac{1}{2}\left( \sigma
_{2}\right) \otimes \left( \sigma _{2}\sigma ^{i}\right) ,  \notag \\
\left\vert 1_{\wQ},0_{\bar{\wQ}};~1_{J}\right\rangle &=&\frac{1}{2}\left( \sigma
_{2}\sigma ^{i}\right) \otimes \left( \sigma _{2}\right) ,  \notag \\
\left\vert 1_{\wQ},1_{\bar{\wQ}};~1_{J}\right\rangle &=&\frac{1}{2\sqrt{2}}%
\varepsilon ^{ijk}\left( \sigma _{2}\sigma ^{j}\right) \otimes \left( \sigma
_{2}\sigma ^{k}\right) .  \label{notations}
\end{eqnarray}%
The properties of these matrices are given in the appendix of ref.~\cite{Maiani:2004vq}.
 The next step is the
diagonalization of the Hamiltonian (\ref{01}) using the basis of states with
definite diquark and antidiquark spin and total angular momentum.
There are two different possibilities~\cite{Maiani:2004vq}: 
Lowest lying $[bq][\bar{b}\bar{q}]$ states $\left( L_{\wQ\bar{\wQ}}=0\right) $ and
higher mass $[bq][\bar{b}\bar{q}]$ states $\left( L_{\wQ\bar{\wQ}}=1\right) $, which we discuss
below.
%
\subsection{Lowest lying $[bq][\bar{b}\bar{q}]$ states $\left( L_{\wQ\bar{\wQ}}=0\right) $}
%
The states can be classified in terms of the diquark and
antidiquark spin, $S_{\wQ}$ and $S_{\bar{\wQ}}$, total angular momentum $J$,
parity, $P$ and charge conjugation, $C$. Considering both
 {\it good} and {\it bad}
diquarks and having $L_{\wQ\bar{\wQ}}=0$ we have six possible states which are
listed below.

\textbf{i. Two states with }$J^{PC}=0^{++}$\textbf{:}%
\begin{eqnarray}
\left\vert 0^{++}\right\rangle &=&\left\vert 0_{\wQ},0_{\bar{\wQ}%
};~0_{J}\right\rangle ;  \notag \\
\left\vert 0^{++\prime }\right\rangle &=&\left\vert 1_{\wQ},1_{\bar{\wQ}%
};~0_{J}\right\rangle .  \label{06}
\end{eqnarray}

\textbf{ii. Three states with }$J=1$\textbf{:}%
\begin{eqnarray}
\left\vert 1^{++}\right\rangle &=&\frac{1}{\sqrt{2}}\left( \left\vert
0_{\wQ},1_{\bar{\wQ}};~1_{J}\right\rangle +\left\vert 1_{\wQ},0_{\bar{\wQ}%
};~1_{J}\right\rangle \right) ;  \notag \\
\left\vert 1^{+-}\right\rangle &=&\frac{1}{\sqrt{2}}\left( \left\vert
0_{\wQ},1_{\bar{\wQ}};~1_{J}\right\rangle -\left\vert 1_{\wQ},0_{\bar{\wQ}%
};~1_{J}\right\rangle \right) ;  \notag \\
\left\vert 1^{+-\prime }\right\rangle &=&\left\vert 1_{\wQ},1_{\bar{\wQ}%
};~1_{J}\right\rangle .  \label{07}
\end{eqnarray}%
All these states have positive parity as both the {\it good} and {\it bad} diquarks
have positive parity and $L_{\wQ\bar{\wQ}}=0$. The difference is in the charge
conjugation quantum number, the state $\left\vert 1^{++}\right\rangle $ is even under
charge conjugation, whereas $\left\vert 1^{+-}\right\rangle $ and $%
\left\vert 1^{+-\prime }\right\rangle $ are odd.

\textbf{iii. One state with }$J^{PC}=2^{++}$\textbf{:}%
\begin{equation}
\left\vert 2^{++}\right\rangle =\left\vert 1_{\wQ},1_{\bar{\wQ}%
};~2_{J}\right\rangle .  \label{08}
\end{equation}

Keeping in view that for $L_{\wQ\bar{\wQ}}=0$ there is no spin-orbit and purely
orbital term, the Hamiltonian (\ref{01}) takes the form%
\begin{eqnarray}
H &=&2m_{[bq]}+2(\mathcal{K}_{bq})_{\bar{3}}[(\mathbf{S}_{b}\cdot \mathbf{S}%
_{q})+(\mathbf{S}_{\bar{b}}\cdot \mathbf{S}_{\bar{q}})]+2\mathcal{K}_{q\bar{q%
}}(\mathbf{S}_{q}\cdot \mathbf{S}_{\bar{q}})  \notag \\
&&+2(\mathcal{K}_{b\bar{q}})(\mathbf{S}_{b}\cdot \mathbf{S}_{\bar{q}}+%
\mathbf{S}_{\bar{b}}\cdot \mathbf{S}_{q})+2\mathcal{K}_{b\bar{b}}(\mathbf{S}%
_{b}\cdot \mathbf{S}_{\bar{b}}).  \label{Haml-zero}
\end{eqnarray}%
The diagonalisation of the Hamiltonian (\ref{Haml-zero}) with the states defined
above gives the eigenvalues which are needed to estimate the masses of these
states. It is straightforward to see that for the $1^{++}$ and $2^{++}$ states
the Hamiltonian is diagonal with the eigenvalues \cite{Maiani:2004vq}%
\begin{eqnarray}
M\left( 1^{++}\right)  &=&2m_{[bq]}-(\mathcal{K}_{bq})_{\bar{3}}+\frac{1}{2}%
\mathcal{K}_{q\bar{q}}-\mathcal{K}_{b\bar{q}}+\frac{1}{2}\mathcal{K}_{b\bar{b%
}},  \label{09} \\
M\left( 2^{++}\right)  &=&2m_{[bq]}+(\mathcal{K}_{bq})_{\bar{3}}+\frac{1}{2}%
\mathcal{K}_{q\bar{q}}+\mathcal{K}_{b\bar{q}}+\frac{1}{2}\mathcal{K}_{b\bar{b%
}}.  \label{10}
\end{eqnarray}%
All other quantities are now specified except the mass of the
constituent diquark. We take the Belle data~\cite{Zupanc:2009qc} as input and
identify the $Y_b(10890)$ with the lightest of the $1^{--}$ states, $Y_{[bq]}$, yielding a
diquark mass  $m_{[bq]}=5.251 \;\tn{GeV}$. This procedure is analogous to what was done
in~\cite{Maiani:2004vq}, in which the mass of the diquark $[cq]$ was fixed by using the mass of
$X(3872)$ as input, yielding $m_{[cq]}=1.933$ GeV. Instead, if we use this determination of
$m_{[cq]}$ and use the formula $m_{[bq]} =m_{[cq]}+\left( m_{b}-m_{c}\right)$, which has the virtue that the
mass difference $m_c-m_b$ is well determined, we get $m_{[bq]}=5.267\;\tn{GeV}$, yielding a difference of
$16 \;\tn{MeV}$. This can be taken as an estimate of the theoretical error on $m_{[bq]}$, which then yields
an uncertainty of about 30 MeV in the estimates of the tetraquark masses of interest for us.

The couplings corresponding to the spin-spin
interactions have been calculated for the colour singlet and colour antitriplet
only. In Eq.~(\ref{02}), however, the quantities  $\mathcal{K}_{q\bar{q}}$, $\mathcal{K}%
_{b\bar{q}}$ and $\mathcal{K}_{b\bar{b}}$ involve both 
colour singlet  and colour octet couplings 
between the quarks and antiquraks in a $\wQ\bar{\wQ}$ system. So for $\mathcal{K}_{b%
\bar{b}}$ \cite{Drenska:2008gr} 
\begin{equation}
\mathcal{K}_{b\bar{b}}\left( [bq][\bar{b}\bar{q}]\right) =\frac{1}{3}\left( 
\mathcal{K}_{b\bar{b}}\right) _{0}+\frac{2}{3}\left( \mathcal{K}_{b\bar{b}%
}\right) _{8}~,  \label{identity4}
\end{equation}%
where $\left( \mathcal{K}_{b\bar{b}}\right) _{0}$ is reported in Table II.
 $\left( \mathcal{K}_{b\bar{b}%
}\right) _{8}$  can be derived from the one gluon exchange model by
using the relation \cite{Maiani:2004vq}:%
\begin{equation}
\left( \mathcal{K}_{b\bar{b}}\right) _{\mathbf{X}}\sim \left( C^{2}\left( 
\mathbf{X}\right) -C^{2}\left( \mathbf{3}\right) -C^{2}\left( \mathbf{\bar{3}%
}\right) \right)~,   \label{identity5}
\end{equation}%
with $C^{2}\left( \mathbf{X}\right) =0$, $4/3$, $4/3$, $3$ for $\mathbf{X=0}$%
, $\mathbf{3}$, $\mathbf{\bar{3}}$, $\mathbf{8}$ respectively. Finally, Eq. (%
\ref{identity4}) gives%
\begin{equation}
\mathcal{K}_{b\bar{b}}\left( [bq][\bar{b}\bar{q}]\right) =\frac{1}{4}\left( 
\mathcal{K}_{b\bar{b}}\right) _{0}~.  \label{identity6}
\end{equation}

Now, we have all the input parameters to calculate the mass spectrum
numerically. Putting everything together the masses for the hidden $b\bar{b}$
tetraquark states $1^{++}$ and $2^{++}$
states are:%
\begin{eqnarray}
M\left( 1^{++}\right)  
&=&10.504\text{ GeV, for }q=u,~d ,  \label{11b} \\
&=&10.849\text{ GeV, for }q=s ,  \label{11b1} \\
&=&13.217\text{ GeV, for }q=c , \label{11ba} \\
M\left( 2^{++}\right)  
&=&10.520\text{ GeV, for }q=u,~d,  \label{11b2} \\
&=&10.901\text{ GeV, for }q=s,  \label{11b3} \\
&=&13.239\text{ GeV, for }q=c.  \label{11bb}
\end{eqnarray}%
For the corresponding $0^{++}$ and $1^{+-}$ tetraquark states, the Hamiltonian is not
 diagonal and
 we have the following $2\times 2$ matrices:%
\begin{equation}
M\left( 0^{++}\right) =\left( 
\begin{array}{cc}
-3(\mathcal{K}_{bq})_{\bar{3}} & \frac{\sqrt{3}}{2}\left( \mathcal{K}_{q\bar{%
q}}+\mathcal{K}_{b\bar{b}}-2\mathcal{K}_{b\bar{q}}\right)  \\ 
\frac{\sqrt{3}}{2}\left( \mathcal{K}_{q\bar{q}}+\mathcal{K}_{b\bar{b}}-2%
\mathcal{K}_{b\bar{q}}\right)  & (\mathcal{K}_{bq})_{\bar{3}}-\left( 
\mathcal{K}_{q\bar{q}}+\mathcal{K}_{b\bar{b}}+2\mathcal{K}_{b\bar{q}}\right) 
\end{array}%
\right) ,  \label{12}
\end{equation}%
\begin{equation}
M\left( 1^{+-}\right) =\left( 
\begin{array}{cc}
-(\mathcal{K}_{bq})_{\bar{3}}+\mathcal{K}_{b\bar{q}}-\frac{\left( \mathcal{K}%
_{q\bar{q}}+\mathcal{K}_{b\bar{b}}\right) }{2} & \mathcal{K}_{q\bar{q}}-%
\mathcal{K}_{b\bar{b}} \\ 
\mathcal{K}_{q\bar{q}}-\mathcal{K}_{b\bar{b}} & (\mathcal{K}_{bq})_{\bar{3}}-%
\mathcal{K}_{b\bar{q}}-\frac{\left( \mathcal{K}_{q\bar{q}}+\mathcal{K}_{b%
\bar{b}}\right) }{2}%
\end{array}%
\right) .  \label{13}
\end{equation}%
To estimate the masses of these two states, one has to diagonalise the above
matrices. After doing this, the mass spectrum of these $b\bar{b}$ states is shown in
Fig. \ref{FullSpectrum}. 
\subsection{Higher mass $[bq][\bar{b}\bar{q}]$ states $\left( L_{\wQ\bar{\wQ}}=1\right) $}
We now discuss orbital excitations with $L_{\wQ\bar{\wQ}}=1$ having both {\it good} and {\it bad}
diquarks. In this paper, we are particularly interested in the $1^{--}$
multiplet. Using the basis vectors defined in reference \cite{Drenska:2008gr}
the mass shift due to the spin-spin interaction terms $H_{SS}$ becomes:%
\begin{equation}
\Delta M_{SS}=\left( 
\begin{array}{ccc}
-3\left( \mathcal{K}_{bq}\right) _{\bar{3}} & 0 & 0 \\ 
0 & -\left( \mathcal{K}_{bq}\right) _{\bar{3}}-\mathcal{K}_{b\bar{q}}+\left( 
\mathcal{K}_{q\bar{q}}+\mathcal{K}_{b\bar{b}}\right) /2 & 0 \\ 
0 & 0 & -\left( \mathcal{K}_{bq}\right) _{\bar{3}}-\mathcal{K}_{b\bar{q}%
}-\left( \mathcal{K}_{q\bar{q}}+\mathcal{K}_{b\bar{b}}\right) /2%
\end{array}%
\right) ~.  \label{16}
\end{equation}%
The eigenvalues of the spin-orbit and angular momentum operators given in Eq. (\ref{01})
 were calculated by Polosa et al.~\cite{Drenska:2008gr}, and we have
summarised these values in Table \ref{tabcoeff}.\footnote{ The entry for $a$ in the last row of
 Table 
 \ref{tabcoeff} differs from the corresponding one
 in the first reference in~\protect\cite{Drenska:2008gr}, which is given as $-2$,
but this point has now been settled  amicably in favour of the value given here.}
\begin{table}[tb]
\caption{Eigenvalues of the spin-orbit and angular momentum operator in Eq. (%
\protect\ref{01}) for the states having $J=L_{\wQ\bar{\wQ}}+S_{\wQ\bar{\wQ}}=1.$%
}
\label{tabcoeff}
\begin{center}
\begin{tabular}{|l|l|l|}
\hline
$\left\vert S_{\wQ}\,\text{, }S_{\bar{\wQ}}\text{, }S_{\wQ\bar{\wQ}}\text{, }%
L_{\wQ\bar{\wQ}} \right\rangle $ & $a\left( S_{\wQ}\,\text{, }S_{\bar{\wQ}}\text{, }S_{\wQ\bar{\wQ}}%
\text{, }L_{\wQ\bar{\wQ}} \right) $ & $b\left( s_{\wQ}\,\text{, }S_{\bar{\wQ}}\text{, }%
S_{\wQ\bar{\wQ}}\text{, }L_{\wQ\bar{\wQ}} \right) $ \\ \hline
$\left\vert 0\,\text{, }0\text{, }0\text{, }1\right\rangle $ & $~~0$ & $1$
\\ \hline
$\left\vert 1\,\text{, }0\text{, }1\text{, }1\right\rangle $ & $-2$ & $1$ \\ 
\hline
$\left\vert 1\,\text{, }1\text{, }2\text{, }1\right\rangle $ & $-6$ & $1$ \\ 
\hline
$\left\vert 1\,\text{, }1\text{, }1\text{, }1\right\rangle $ & $-2$ & $1$ \\ 
\hline
$\left\vert 1\,\text{, }1\text{, }0\text{, }1\right\rangle $ & $~~0$ & $1$
\\ \hline\hline
\end{tabular}%
\end{center}
\end{table}

Hence the eight tetraquark states $[bq][\bar{b}\bar{q}]$ ($q=u, d$) having the quantum numbers
 $1^{--}$ are:%
\begin{eqnarray}
M_{Y_{[bq]}}^{(1)}\left( S_{\wQ}=0,~S_{\bar{\wQ}}=0,~S_{\wQ\bar{\wQ}}=0,~L_{\wQ\bar{\wQ}}=1\right) 
&=&2m_{\left[ bq\right] }+\lambda _{1}+B_{\mathcal{\wQ}},  \notag \\
M_{Y_{[bq]}}^{(2)}\left( S_{\wQ}=1,~S_{\bar{\wQ}}=0,~S_{\wQ\bar{\wQ}}=1,~L_{\wQ\bar{\wQ}}=1\right) 
&=&2m_{\left[ bq\right] }+\Delta +\lambda _{2}-2A_{\wQ}+B_{\wQ}, 
\notag \\
M_{Y_{[bq]}}^{(3)}\left( S_{\wQ}=1,~S_{\bar{\wQ}}=1,~S_{\wQ\bar{\wQ}}=0,~L_{\wQ\bar{\wQ}}=1\right) 
&=&2m_{\left[ bq\right] }+2\Delta +\lambda _{3}+B_{\wQ},  \label{17}
\\
M_{Y_{[bq]}}^{(4)}\left( S_{\wQ}=1,~S_{\bar{\wQ}}=1,~S_{\wQ\bar{\wQ}}=2,~L_{\wQ\bar{\wQ}}=1\right) 
&=&2m_{\left[ bq\right] }+2\Delta +\lambda _{3}-6A_{\wQ}+B_{\wQ}, 
\notag
\end{eqnarray}%
where $\lambda_i(i=1,2,3) $ are the diagonal elements of the matrix $\Delta M_{SS}$ given in
Eq.~(\ref{16}). Note that there are 16 electrically neutral self-conjugate $1^{--}$ tetraquark
 states $Y_{[bq]}^{(n)}$
with the quark contents $[bq][\bar{b}\bar{q}]$, with $q=u,d,s$ or $c$, of which the two corresponding to
$[bu][\bar{b}\bar{u}]$ and  $[bd][\bar{b}\bar{d}]$, i.e., $Y_{[bu]}^{(n)}$
and $Y_{[bd]}^{(n)}$ are degenerate in mass due to the
isospin symmetry. There are yet more electrically neutral $J^{PC}=1^{--}$ states with the 
mixed light quark content $[bd][\bar{b}\bar{s}]$ and their charge conjugates
$[bs][\bar{b}\bar{d}]$. However, 
these mixed states don't couple directly to the photons, $Z^0$ or the gluon, and are not of
immediate interest to us in this paper.

The numerical values of the coefficients corresponding to 
$A_{\wQ}$ and $B_{\wQ}$ are given in Table \ref{tabcoeff} and are
labelled by $a$ and $b$, respectively. The quantity $\Delta $ is the mass difference
of the {\it good} and the {\it bad} diquarks, i.e.%
\begin{equation}
\Delta =m_{\wQ}\left( S_{\wQ}=1\right) -m_{\wQ}\left( S_{\wQ}=0\right) .  \label{18}
\end{equation}%
In order to calculate the numerical values of these states, we have to
estimate $\Delta $ which is the only
unknown remaining in this calculation. Following Jaffe and Wilczek
\cite{Jaffe:2004ph}, the value of $\Delta $ for diquark $[bq]$ is
 $\Delta =202$
MeV for $q=u$, $d$, $s$ and $c$ quarks. We recall that we have used the
known mesons and baryons to calculate the couplings of the spin-spin
interaction and we can extend the same procedure to the $S=1$, $L=(0$, $1$)
meson states $B^{\ast }$, $B_{1}\left( 5721\right) $, $B_{2}\left(
5747\right) $ to calculate the values of $A_{\wQ}$ and $B_{\mathcal{Q}}$ which
are:%
\begin{eqnarray}
A_{\wQ} &=&5\text{ MeV, for }q=u\text{, }d,  \notag \\
A_{\wQ} &=&3\text{ MeV, for }q=s\text{, }c , \notag \\
B_{\wQ} &=&408\text{ MeV, for }q=u\text{, }d,  \notag \\
B_{\wQ} &=&423\text{ MeV, for }q=s,c . \label{19}
\end{eqnarray}%
Numerical values of the masses for the
states given in Eq.~(\ref{17}) are quoted in Table V. Some of the entries,
in particular $M_{{Y}_{[bq]}^{(1)}}$ ($ q=u,d,s$),  
are comparable with the existing ones in 
 refs.~\cite{Ebert:2008se,Wang:2009kw}.

\begin{figure}[h!]
\centering
\includegraphics[width=0.4\textwidth]{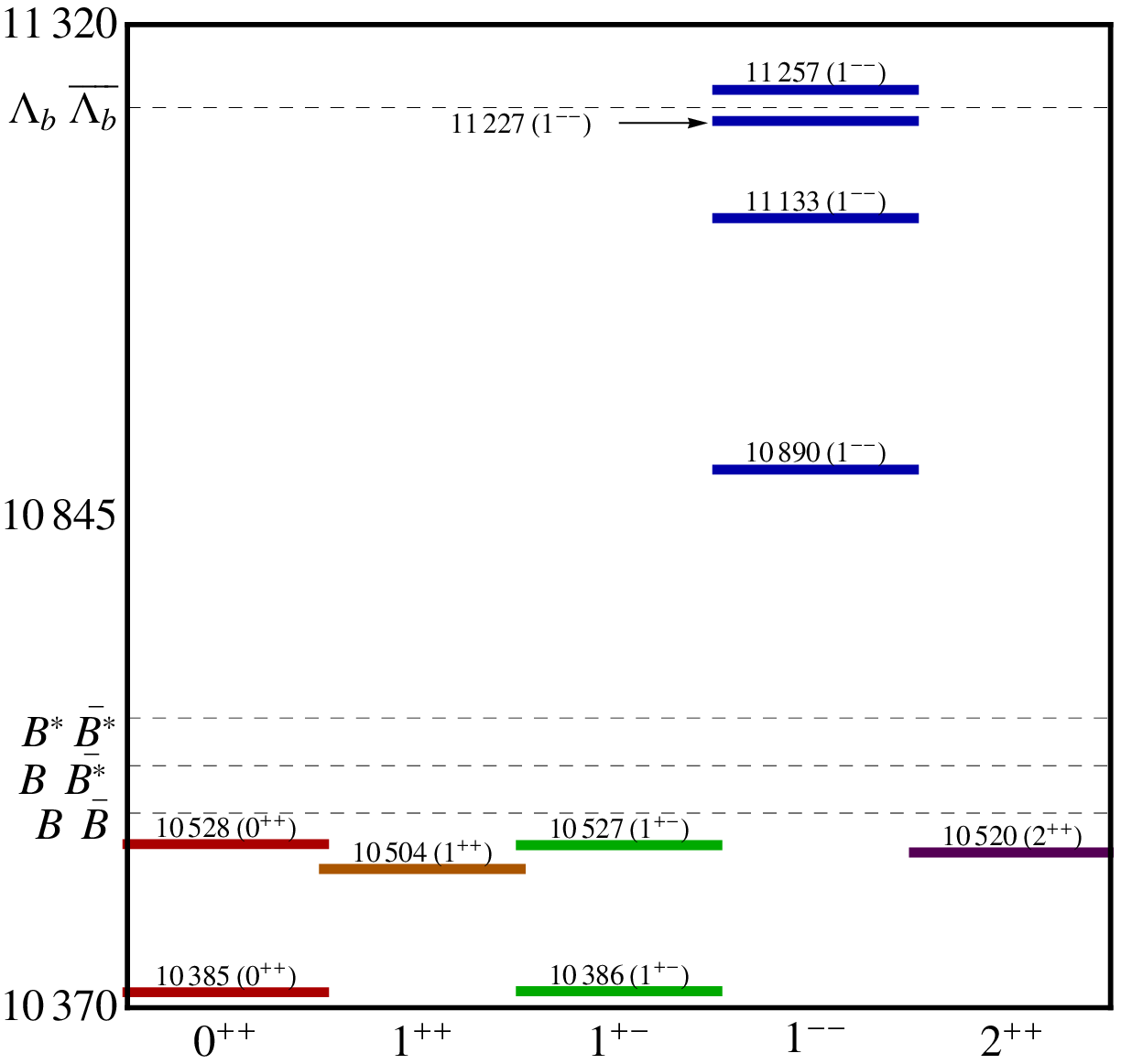} \enspace\enspace%
\enspace
\includegraphics[width=0.4\textwidth]{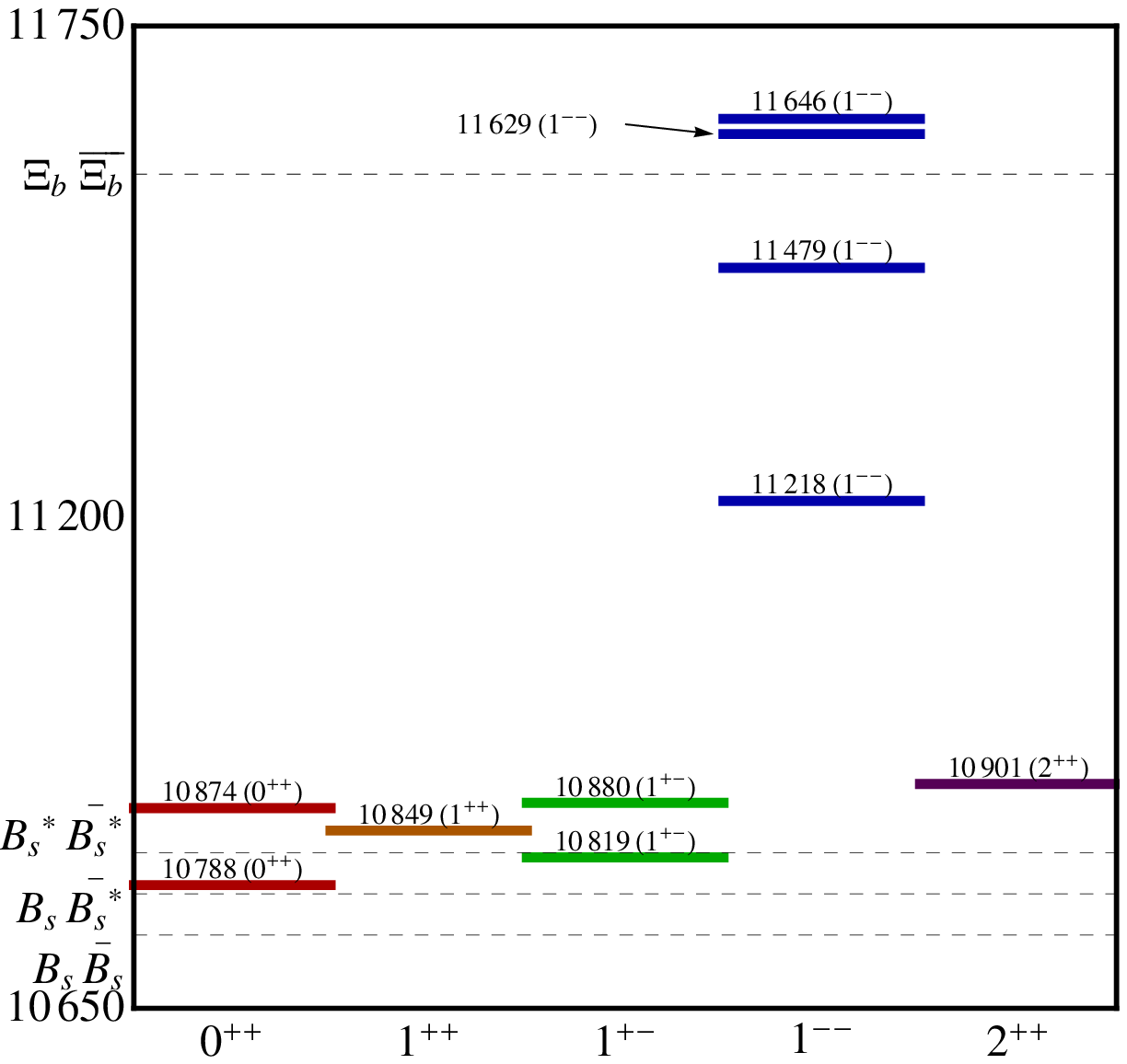} %
\includegraphics[width=0.4\textwidth]{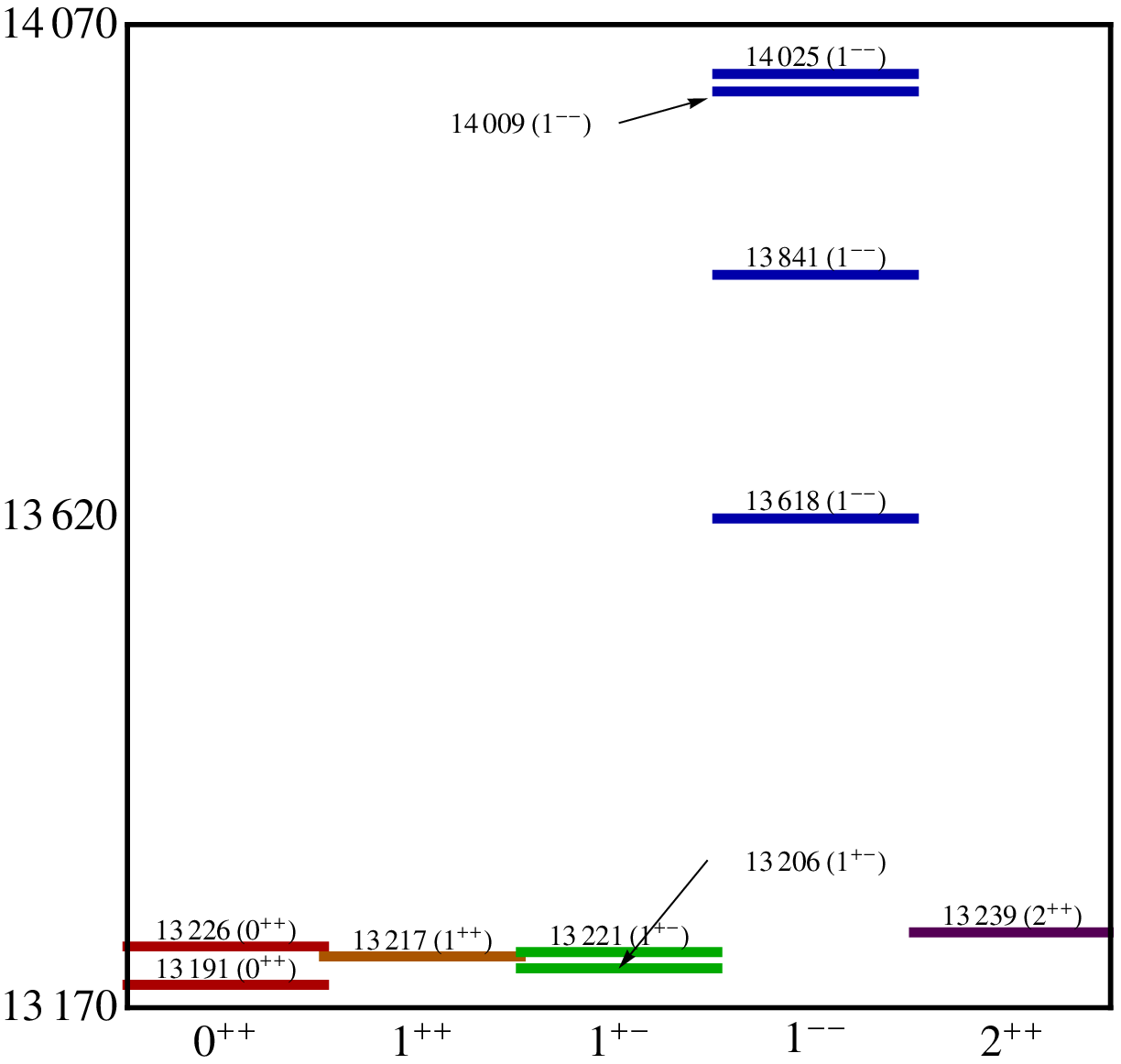} \enspace\enspace%
\enspace
\includegraphics[width=0.4\textwidth]{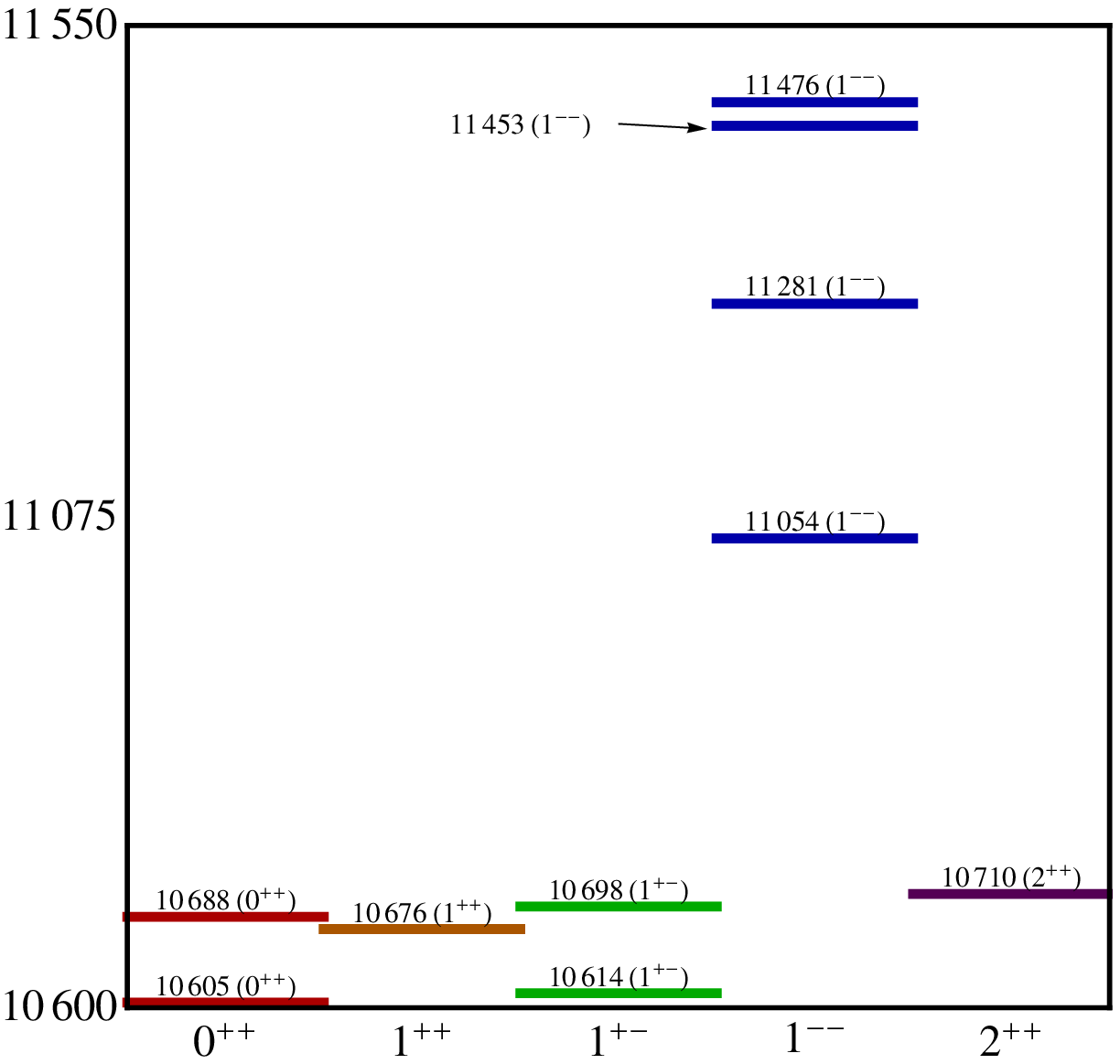}
\caption{Tetraquark mass spectrum with the valence quark
content $[bq][\bar{b}\bar{q}]$ with $q=u,d$, assuming isospin symmetry (upper left frame), with $q=s$
(upper right frame), with $q=c$ (lower left frame), and for the mixed light quark content
$[bd][\bar{b}\bar{s}]$ (lower right frame).  
Some important decay thresholds are indicated by dashed lines. The value 10890 is an input
 for the
lowest $J^{PC}=1^{--}$ tetraquark state $Y_{[bq]}^{(1)}$. All masses are given in MeV. }
\label{FullSpectrum}
\end{figure}

 Finally, the mass spectrum for the tetraquark states $[bq][\bar{b} \bar{q}]$
for $q=u,d,s,c$ with $J^{PC}=0^{++}, 1^{++}, 1^{+-}, 1^{--}$ and $2^{++}$
 states is plotted in Fig. \ref%
{FullSpectrum} in the isospin-symmetry limit. The $b \bar{b}$ tetraquark states with mixed 
light quark content $[bd][\bar{b}\bar{s}]$ are also shown in this figure.
Of these, the $1^{--}$ state $Y^{(1)}_{[bq]}(\BBConstmassY)$  shown in the upper left
frame in Fig. \ref{FullSpectrum} is of central interest to us in this paper. 

\begin{table}[tb]
\caption{Masses of the $1^{--}$ tetraquark states $M_{{Y}_{[bq]}}^{(n)}$  in GeV 
 as computed from
Eqs. (\protect\ref{17}), (\protect\ref{18}) and (\protect\ref{19}). The value  
$M_{{Y}_{[bq]}^{(1)}}$  (for $q=u,d$) is fixed to be 10.890 GeV, identifying this with the
mass of the $Y_b$ from Belle~\protect\cite{Zupanc:2009qc}}.
\begin{center}
\begin{tabular}{|l|l|l|l|l|}
\hline
$M_{{Y}_{[bq]}}^{(i)}$ & $q=u$, $d$ & $q=s$ & $q=c$ & $q=d$, $\bar{q}=\bar{s}$ \\ \hline
$M_{{Y}_{[bq]}^{(1)}}$ & $\BBConstmassY$ & $11.218$ & $13.618$ & $11.054$ \\ \hline
$M_{{Y}_{[bq]}^{(2)}}$ & $\BBConstmassYTwo$ & $11.479$ & $13.841$ & $11.281$ \\ \hline
$M_{{Y}_{[bq]}^{(3)}}$ & $11.257$ & $11.646$ & $14.025$ & $11.476$ \\ \hline
$M_{{Y}_{[bq]}}^{(4)}$ & $11.227$ & $11.629$ & $14.009$ & $11.453$ \\ \hline\hline
\end{tabular}%
\end{center}
\end{table}
\section{Isospin Breaking and Leptonic Decay Widths of the $J^{PC}=1^{--}$ Tetraquarks}
We discuss in this section the isospin breaking effects, which were neglected in the previous section,
and calculate the decay widths $\Gamma_{ee}(Y_{[b,l]})$ and $\Gamma_{ee}(Y_{[b,h]})$
for $Y_{[b, l]}$ and $Y_{[b, h]}$.
The mass eigenstates are given by a linear superposition of the states defined in
 \eqref{def_high_low_states}. Introducing a mixing angle $\theta$, we have, for the lighter and
 heavier states: 
\begin{eqnarray}
\label{low_high_def}
Y_{[b,l]}&=&\cos \theta\;  Y_{[bu]} + \sin \theta \;Y_{[bd]}, \\
Y_{[b,h]}&=&-\sin \theta  \;Y_{[bu]} + \cos \theta \;Y_{bd]}.
\end{eqnarray}
The isospin breaking part of the mass matrix is 
\begin{equation}
\left( 
\begin{array}{cc}
2 m_u +\delta & \delta \\
\delta & 2 m_d +\delta
\end{array}
\right),
\end{equation}
where $\delta$ is the contribution from quark annihilation diagrams, where the light quark pair
 annihilates to intermediate gluons. Taking this into account,
the isospin mass breaking is given by\footnote{The expression (\ref{isomassbreak}) differs from the one derived in 
 \cite{Maiani:2004vq}, but there is consensus now on the expression given here.} 
\begin{equation}\label{isomassbreak}
M(Y_{[b,h]})-M(Y_{[b,l]})=(7\pm 3) \; \cos(2\theta)\;{\rm MeV}.
\end{equation}
 
 The partial electronic widths $\Gamma _{ee}(Y_{[b,l]})$
and $\Gamma _{ee}(Y_{[b,h]})$ are  given by the well-known Van Royen-Weisskopf formula for the
 P-states, which we write generically as: 
\begin{equation}
\Gamma _{ee}=\frac{16\pi Q^{2}\alpha ^{2}|\Psi_{\wQ \bar{\wQ}}^{\prime} (0)|^2}
{M^{2}\omega ^{2}},
\end{equation}
where $Q=Q_{[bd]}= -2/3$ is the diquark charge in $Y_{bd}=[bd][\bar{b}\bar{d}]$ and
 $Q=Q_{[bu]}= +1/3$ is the charge of the
 diquarks in $Y_{bu}=[bu][\bar{b}\bar{u}]$, 
 and $\Psi_{\wQ \bar{\wQ}}^{\prime }(\vec{r})=\psi (\phi ,\theta)R^{\prime }(r)$ is the first derivative in $r$ of the wave function of the tetraquark, which needs to be taken at the origin, i.e. $\Psi_{\wQ \bar{\wQ}}^{\prime }(0)=\sqrt{3/(4 \pi)}R^{\prime }(0)$.
We have approximated $\omega$ by the diquark mass.

We determine the
wave functions for the P-state tetraquarks $[bd][\bar{b}\bar{d}]$ and $[bu][\bar{b}\bar{u}]$
from the corresponding wave functions for the P-state $b\bar{b}$ system by scaling the string tension
in the linear part of the potential, as discussed in the introduction.
As most potential models agree in their linear (confining) parts~\cite{Buchmuller:1980su} and
the linear part of the  potential essentially determines the heavy Quarkonia wave functions, the
uncertainty in $\Psi_{b\bar{b}}(0)$ from the underlying model is not a concern. We have used 
the  QQ-onia package of~\cite{DomenechGarret:2008sr}, yielding
$\vert R^{\prime }(0)\vert^2= 2.062 \enspace GeV^{5}$ for the $b\bar{b}$ radial
wave function, which we have used as normalisation. The corresponding value for
the tetraquark states $[bq][\bar{b}\bar{q}]$ is then calculated as
 $\Psi_{\wQ \bar{\wQ}}(0) \simeq \kappa \Psi_{b\bar{b}}(0)$, and used in our derivations of the decay widths.
 We expect that for all the
P-states $Y^{(n)}_{[bu]}$ and $Y^{(n)}_{[bd]}$, the electronic widths will be constant, to a
good approximation.

The ratio ${\mathcal{R}}_{ee}(Y_b)$ of $\Gamma_{ee}(Y_{[b,l]})$ and $\Gamma_{ee}(Y_{[b,h]})$
 is given by 
\begin{equation}\label{ratiodef}
{\mathcal{R}}_{ee}(Y_b) \equiv \frac{\Gamma_{ee}(Y_{[b,l]})}{\Gamma_{ee}(Y_{[b,h]})}=
\frac{Q_l^2(\theta)}{Q_h^2(\theta)}
=\left[ \frac{1-2\tan\theta}{2+ \tan\theta}\right]^2~,
\end{equation}
where $Q_l(\theta)=Q_{[bu]} \cos \theta + Q_{[bd]}\sin \theta$ and
$Q_h(\theta)= -Q_{[bu]} \sin \theta + Q_{[bd]} \cos \theta$ are the mixing-angle weighted 
charges.
Since the total cross sections for
 $e^+ e^- \to (Y_{[b,l]},Y_{[b,h]}) \to {\rm hadrons}$ are directly proportional
 to $\Gamma_{ee}(Y_{[b,l]})$ and $\Gamma_{ee}(Y_{[b,h]})$, the ratio 
 ${\mathcal{R}}_{ee}(Y_b)$ is accessible from the experiment. The absolute values of the
decay widths $\Gamma_{ee}(Y_{[b,l]})$
and $\Gamma_{ee}(Y_{[b,h]})$ are given by
 $\Gamma_{ee}(Y_{[b,i]})=0.4\; \kappa^2 Q_i(\theta)^2$ keV, where $Q_i(\theta)$ are the mixing angle
weighted charges of the two mass eigenstates, $Y_{[b,l]}$ and $Y_{[b,h]}$. 
which can also be seen  in \eqref{ratiodef}.

\section{Diquark-antidiquark decay modes}
In this section we discuss the dominant hadronic decays of the $L_{\wQ\bar{\wQ}}=1$ states. In doing this, we restrict
ourselves to the two-body decays, $Y_{[bq]} \to B_q^{(*)} \bar{B}_q^{(*)}$, and when allowed kinematically,
also the decay  $Y_{[bq]} \to \Lambda_b \bar{\Lambda}_b$. Their thresholds are pictured in figure \ref{FullSpectrum}.
 These decays are Zweig allowed and involve essentially
quark rearrangements and the possible pop-up of a light $q\bar{q}$ pair to make the $\Lambda_b \bar{\Lambda}_b$
state.  
The decays $Y_{[bq]}\to \Upsilon(1S, 2S) \; \pi^+\pi^-$ are also Zweig allowed. However, they
are sub-dominant  and can be neglected in estimating the total decay widths. 

  The vertices and the corresponding decay widths  of the dominant decays are given below:

\begin{equation}  \label{verticesdef1}
\begin{large}
\begin{array}{rclrcl}
\raisebox{-30pt}{\includegraphics[width=0.30\textwidth]{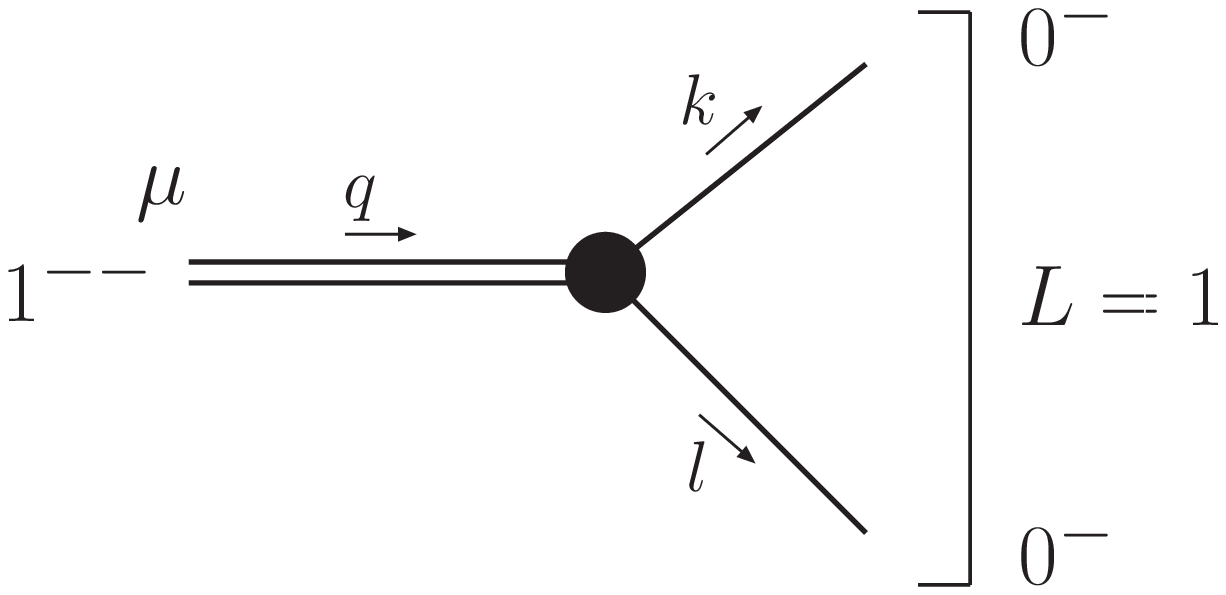}} 
& \widehat{=} & F(k^{\mu}-l^{\mu}) & 
\Longrightarrow\;\;\Gamma & = & \frac{F^2 |\vec{k}|^3}{2 M^2 \pi}, \\ 
\raisebox{-30pt}{\includegraphics[width=0.27\textwidth]{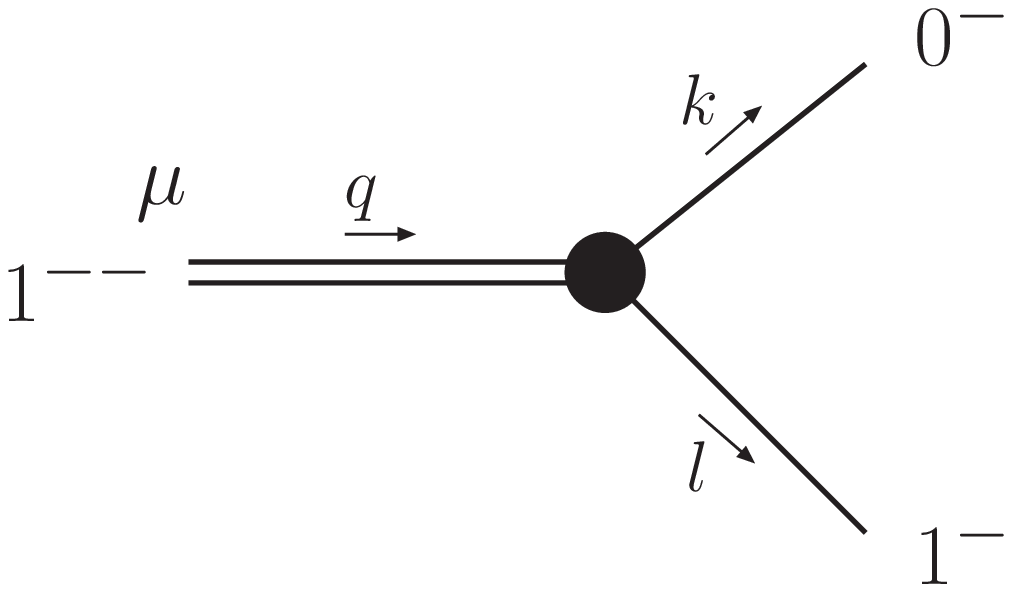}} 
& \widehat{=} & \frac{F }{M}%
\epsilon^{\mu\nu\rho\sigma} k_{\rho}l_{\sigma} & \Longrightarrow\;\;\Gamma & = & \frac{F^2 |%
\vec{k}|^3}{4 M^2 \pi}, \\ 
\raisebox{-30pt}{\includegraphics[width=0.30\textwidth]{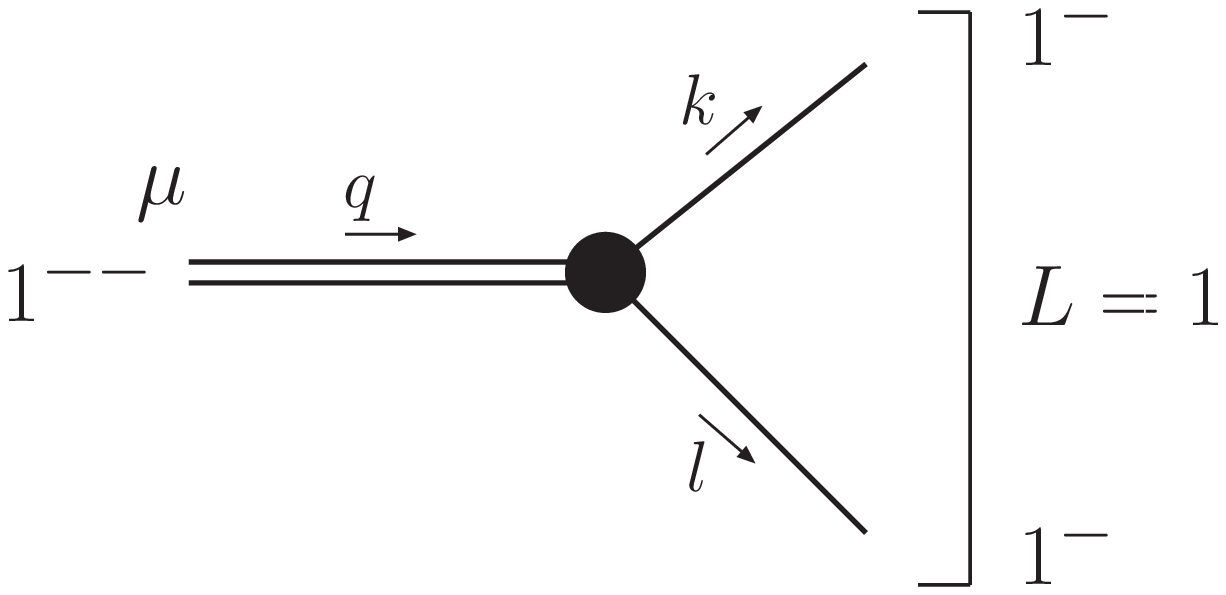}} 
& \widehat{=} & 
\begin{array}{l}
F( g^{\mu\rho}(q+l)^{\nu} \\ 
\enspace\enspace -g^{\mu\nu}(k+q)^{\rho} \\ 
\enspace\enspace +g^{\rho\nu}(q+k)^{\mu} )%
\end{array}
& \Longrightarrow\;\;\Gamma & = & \frac{F^2 |\vec{k}|^3 (48 |\vec{k}|^4-104 M^2|\vec{k}|^2+27
M^4 )}{2\pi(M^3-4|\vec{k}|^2 M)^2},\\
\raisebox{-30pt}{\includegraphics[width=0.27\textwidth]{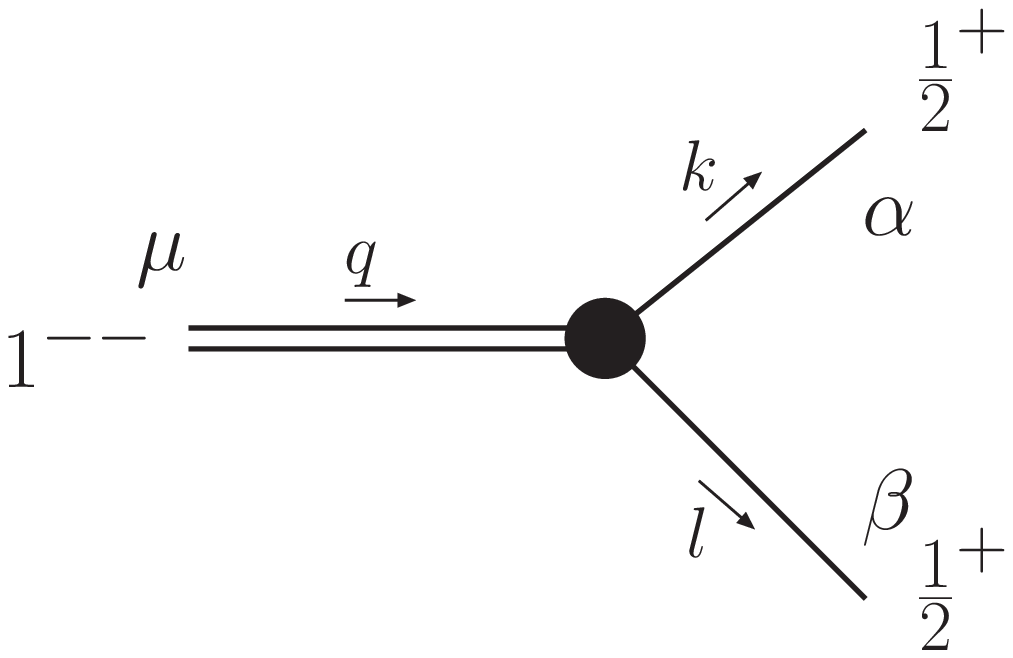}} 
& \widehat{=} & 
\left(F\gamma^{\mu}+\frac{2 F^{\prime }}{i M} q_{\nu}\sigma^{\nu\mu}%
\right)_{\alpha\beta} & \Longrightarrow\;\;\Gamma & = & \frac{3(F^2 +F^{\prime 2})|\vec{k}|}{4
\pi}-\frac{(F^2 + 2 F^{\prime 2})|\vec{k}|^3}{ M^2 \pi} .
\end{array}%
\end{large}
\end{equation}
The centre-of-mass momentum $|\vec{k}|$ is given by 
\begin{equation}
|\vec{k}|=\frac{\sqrt{M^2-(M_1+M_2)^2} \sqrt{M^2-(M_1-M_2)^2} }{2M},
\end{equation}
where $M$ is the mass of the decaying particle and $M_1$, $M_2$ are the masses of the
 decay products. The matrix elements are obtained by multiplying the vertices in 
\eqref{verticesdef1} by the polarisation vectors. Thus, for the decay 
 $Y_{[b,q]} \to B_q \bar{B}_q$, the Lorentz-invariant matrix element is  given by
 $\mathcal{M}=\varepsilon^{Y_{[b,q]}}_{\mu}F(k^{\mu}-l^{\mu})$, and likewise for the other decays shown above.
The decay constants $F$ and $F^{\prime }$ are non-perturbative quantities. We estimate them using the known two-body
decays of $\Upsilon(5S)$, which are described by the same vertices as given above. The different hadronic sizes of the
$b\bar{b}$ Onia states and the tetraquarks $Y_{[bq]}$ are taken into account by the quantity $\kappa$, discussed
 earlier. We use the partial decay widths for the decays
 $\Upsilon (5S) \to B \bar{B}, B \bar{B}^*, B^* \bar{B}^*$ 
from the PDG values of the full width, given as
 $\Gamma_{\rm tot}[\Upsilon(5S)]=
110 \pm 13$ MeV~\cite{Amsler:2008zzb} and the respective branching ratios.
 They are called $\Gamma_{\rm PDG}$ and
given in Table~\ref{mimicprocess}, yielding the coupling constants, called $F_{\rm PDG}$,
and $|\vec{k}|$.
 For the decays $Y^{(i)}_{[bq]} \to \Lambda_b \bar{\Lambda}_b$ and $Y^{(i)}_{[bs]} \to \Xi \bar{\Xi}$, we take
 $F=F'=1.1^{+0.3}_{-0.35}$, and include a factor of $1/3$ for the baryonic final state 
to take into account the  creation of the $q\bar{q}$ pair from the vacuum. We remark that the
estimates of $F_{\rm PDG}$ will be modified, if as anticipated by the BaBar $R_b$-analysis~\cite{:2008hx},
 the decay width $\Gamma_{\rm tot}[\Upsilon(5S)]$ has a significantly lower value. 

 The input values for the 
masses used in our calculation are listed in Table~\ref{tabmassesused}. With this input, our estimates of the
 decay widths for $Y^{(i)}_{[bq]}$ are given  in Table~\ref{estimatesforGamma}. We also give the total
decay widths (up to the factor $\kappa^2$).  As seen in this table, the lowest lying $1^{--}$ states $Y^{(1)}_{[bq]}$
 are expected to have decay widths of $O(50)$ MeV, for $\kappa^2=0.5$. Thus,  the decay
widths of $Y^{(1)}_{[bq]}$ are consistent with the corresponding measurements by Belle,
if we identify $Y^{(1)}_{[bq]}$ with their $Y_b$. The higher $1^{--}$ states have
much larger decay widths and will be correspondingly more difficult to find.  
%
%
\begin{table}[h!]
\caption{Input masses taken from \protect\cite{Amsler:2008zzb} in units of GeV. }
\label{tabmassesused}
\begin{center}
\begin{tabular}{|l|l||l|l||l|l|}
\hline
hadron & mass & hadron & mass & hadron & mass  \\ \hline
$B $ & 5.279 & $\pi$ & 0.139 & $\Upsilon(1S)$ & 9.46  \\ \hline
$B^* $ & 5.325  & $\Lambda_b$ & 5.62 & $\Upsilon(4S) $ & 10.5794 \\ \hline
$B_s $ & 5.366 & $\Xi_b$ & 5.792 & $\Upsilon(10860)$ & 10.865 \\ \hline
$B_s^*$ & 5.412 & $K $ & 0.4937  & $\Upsilon(11020)$ & 11.019     \\ 
\hline\hline
\end{tabular}%
\end{center}
\end{table}
\begin{table}[h!]
\caption{2-body decays $\Upsilon(5S) \to B^{(*)} \bar{B}^{(*)}$, which we use as a reference,
with the mass and the decay widths taken from \protect\cite{Amsler:2008zzb}. The extracted values
of the coupling constants $F_{\rm PDG}$ and the centre of mass momentum $|\vec{k}|$ are also shown.}
\label{mimicprocess}
\begin{center}
\begin{tabular}{|l|l|l|l|}
\hline
process & $\Gamma_{\rm PDG}[ {\rm MeV} ] $ & $F_{\rm PDG}  $ & $|\vec{k}| [ {\rm GeV} ]$ \\ \hline
$\Upsilon(10860)\rightarrow B\enspace \bar{B} $ & $<13.2 $ & $<2.15 $ & $1.3 $ \\ \hline
$\Upsilon(10860)\rightarrow B\enspace \bar{B}^* $ & $15.4^{+6.6}_{-6.6} $ & $3.7^{+0.7}_{-0.9} $ & $1.2 $ \\ \hline
$\Upsilon(10860)\rightarrow B^*\enspace \bar{B}^* $ & $48^{+11}_{-11} $ & $1^{+0.13}_{-0.12} $ & $1.0 $ \\ \hline
\end{tabular}
\end{center}
\end{table}
\begin{table}[h!]
\caption{ Reduced partial decay widths for the tetraquarks $Y^{(i)}_{[bq]}$, the extracted value of the 
coupling constant $F$ and the centre of mass momentum  $|\vec{k}|$ (top left). The reduced total decay widths for
$Y^{(i)}_{[bq]}$ are also tabulated (top right) and for the tetraquarks $Y^{(i)}_{[bs]}$ (the lower two tables).
  The errors in the entries correspond to the errors in the decay widths in Table \protect\ref{mimicprocess}.
  }
\label{estimatesforGamma}
\begin{center}
\begin{tabular}{|l|l|l|l|}
\hline
\text{Decay Mode}                                           & $\Gamma/\kappa^2 [ \tn{MeV} ]$& $F $                   & $|\vec{k}| [ \tn{GeV} ] $ \\ \hline
$Y_{[bq]}^{(1)}\rightarrow B\enspace \bar{B} $                  & $<15 $          & $2.15 $                   & $1.3 $ \\ \hline
$Y_{[bq]}^{(1)}\rightarrow B\enspace \bar{B}^*     $                  & $18_{-8}^{+8} $ & $3.7 $                 & $1.2 $ \\ \hline
$Y_{[bq]}^{(1)}\rightarrow B^*\enspace \bar{B}^* $              & $56_{-14}^{+14}$& $1 $                 & $1.1 $ \\ \hline
$Y_{[bq]}^{(2)}\rightarrow B\enspace \bar{B} $                  & $<33 $          & $2.15 $                 & $1.8 $ \\ \hline
$Y_{[bq]}^{(2)}\rightarrow B\enspace \bar{B}^* $                      & $43_{-18}^{+18} $ & $3.7 $                 & $1.7 $ \\ \hline
$Y_{[bq]}^{(2)}\rightarrow B^*\enspace \bar{B}^* $              & $162_{-42}^{+42}$& $1 $                & $1.6 $ \\ \hline
$Y_{[bq]}^{(3)}\rightarrow B\enspace \bar{B} $                  & $<43 $          & $2.15 $                 & $2 $   \\ \hline
$Y_{[bq]}^{(3)}\rightarrow B\enspace \bar{B}^* $                      &$58_{-25}^{+25} $& $3.7 $                 & $1.9 $ \\ \hline
$Y_{[bq]}^{(3)}\rightarrow B^*\enspace \bar{B}^* $              &$231_{-60}^{+60}$ & $1 $                & $1.8 $ \\ \hline
$Y_{[bq]}^{(3)}\rightarrow \Lambda_b \enspace \bar{\Lambda}_b $ & $10_{-5}^{+5}$& $1.1^{+0.3}_{-0.35}/3 $& $0.3 $ \\ \hline
$Y_{[bq]}^{(4)}\rightarrow B\enspace \bar{B} $                  & $<41 $          & $2.15 $                 & $1.9 $   \\ \hline
$Y_{[bq]}^{(4)}\rightarrow B\enspace \bar{B}^* $                      & $54_{-23}^{+23}$& $3.7 $                 & $1.8 $ \\ \hline
$Y_{[bq]}^{(4)}\rightarrow B^*\enspace \bar{B}^* $              & $213_{-55}^{+55}$& $1 $                & $1.8 $ \\ \hline
\hline
\end{tabular}%
\enspace\enspace\enspace%
\begin{tabular}{|l|l|}
\hline
$1^{--}$ Tetraquark  & $\Gamma_{tot}/\kappa^2 [ \tn{MeV} ] $ \\ \hline
$Y_{[bq]}^{(1)}$ & $88 \pm 16$     \\ \hline
$Y_{[bq]}^{(2)}$ & $238 \pm 48$     \\ \hline
$Y_{[bq]}^{(3)}$ & $342 \pm 65 $     \\ \hline
$Y_{[bq]}^{(4)}$ & $308 \pm 60 $     \\ \hline\hline
\end{tabular}%
\\[0pt]
\vspace{1cm}%
\begin{tabular}{|l|l|l|l|}
\hline
Decay Mode \enspace\enspace\enspace\enspace\enspace\enspace\enspace\enspace%
\enspace\enspace\enspace\enspace\enspace\enspace    & $\Gamma/\kappa^2 [ \tn{MeV} ] $ & $F%
\enspace\enspace\enspace\enspace\enspace$           & $|\vec{k}| [ \tn{GeV} ] $ \\ \hline
$Y_{[bs]}^{(1)}\rightarrow B_s\enspace \bar{B}_s $       & $< 26 $             & $2.15 $                 & $1.6 $ \\ \hline
$Y_{[bs]}^{(1)}\rightarrow B_s\enspace \bar{B}^*_s $           & $33_{-14}^{+14} $   & $3.7 $                  & $1.6 $\\ \hline
$Y_{[bs]}^{(1)}\rightarrow B^*_s\enspace \bar{B}^*_s $   & $118_{-30}^{+30} $  & $1 $                    & $1.5 $ \\ \hline
$Y_{[bs]}^{(2)}\rightarrow B_s\enspace \bar{B}_s $       & $< 47 $             & $2.15 $                 & $2 $ \\ \hline
$Y_{[bs]}^{(2)}\rightarrow B_s\enspace \bar{B}^*_s $           & $64_{-27}^{+27} $   & $3.7 $                  & $2 $\\ \hline
$Y_{[bs]}^{(2)}\rightarrow B^*_s\enspace \bar{B}^*_s $   & $258_{-65}^{+65} $  & $1 $                    & $1.9 $ \\ \hline
$Y_{[bs]}^{(3)}\rightarrow B_s\enspace \bar{B}_s $       & $< 63 $             & $2.15 $                 & $2.3 $\\ \hline
$Y_{[bs]}^{(3)}\rightarrow B_s\enspace \bar{B}^*_s $           & $86_{-37}^{+37} $   & $3.7 $                  & $2.2 $\\ \hline
$Y_{[bs]}^{(3)}\rightarrow B^*_s\enspace \bar{B}^*_s $   & $367_{-90}^{+90} $& $1 $                    & $2.1 $ \\ \hline
$Y_{[bs]}^{(3)}\rightarrow \Xi \enspace \bar{\Xi} $      & $19_{-10}^{+10} $   & $1.1^{+0.3}_{-0.35}/3 $ & $0.6 $ \\ \hline
$Y_{[bs]}^{(4)}\rightarrow B_s\enspace \bar{B}_s $       & $< 61 $             & $2.15 $                 & $2.2 $\\ \hline
$Y_{[bs]}^{(4)}\rightarrow B_s\enspace \bar{B}^*_s $           & $84_{-35}^{+35} $   & $3.7 $                  & $2.2 $\\ \hline
$Y_{[bs]}^{(4)}\rightarrow B^*_s\enspace \bar{B}^*_s $   & $355_{-90}^{+90} $& $1 $                    & $2.1 $ \\ \hline
$Y_{[bs]}^{(4)}\rightarrow \Xi \enspace \bar{\Xi} $      & $16_{-10}^{+10} $   & $1.1^{+0.3}_{-0.35}/3 $ & $0.5 $ \\ \hline
\hline
\end{tabular}%
\enspace\enspace\enspace%
\begin{tabular}{|l|l|}
\hline
 $1^{--}$ Tetraquark & $\Gamma_{tot}/\kappa^2 [ \tn{MeV} ] $ \\ \hline
$Y_{[bs]}^{(1)}$ & $176 \pm 33$ \\ \hline
$Y_{[bs]}^{(2)}$ & $368 \pm 70 $ \\ \hline
$Y_{[bs]}^{(3)}$ & $534 \pm 100 $ \\ \hline
$Y_{[bs]}^{(4)}$ & $516 \pm 96 $ \\ \hline\hline
\end{tabular}%
\end{center}
\end{table}
\section{Analysis of the BaBar $R_b$ energy scan and possible signal  of a $b\bar{b}$
 tetraquark state at  \BBfitmassY~GeV}
\begin{figure}[H]
\centering
\includegraphics[width=0.45\textwidth]{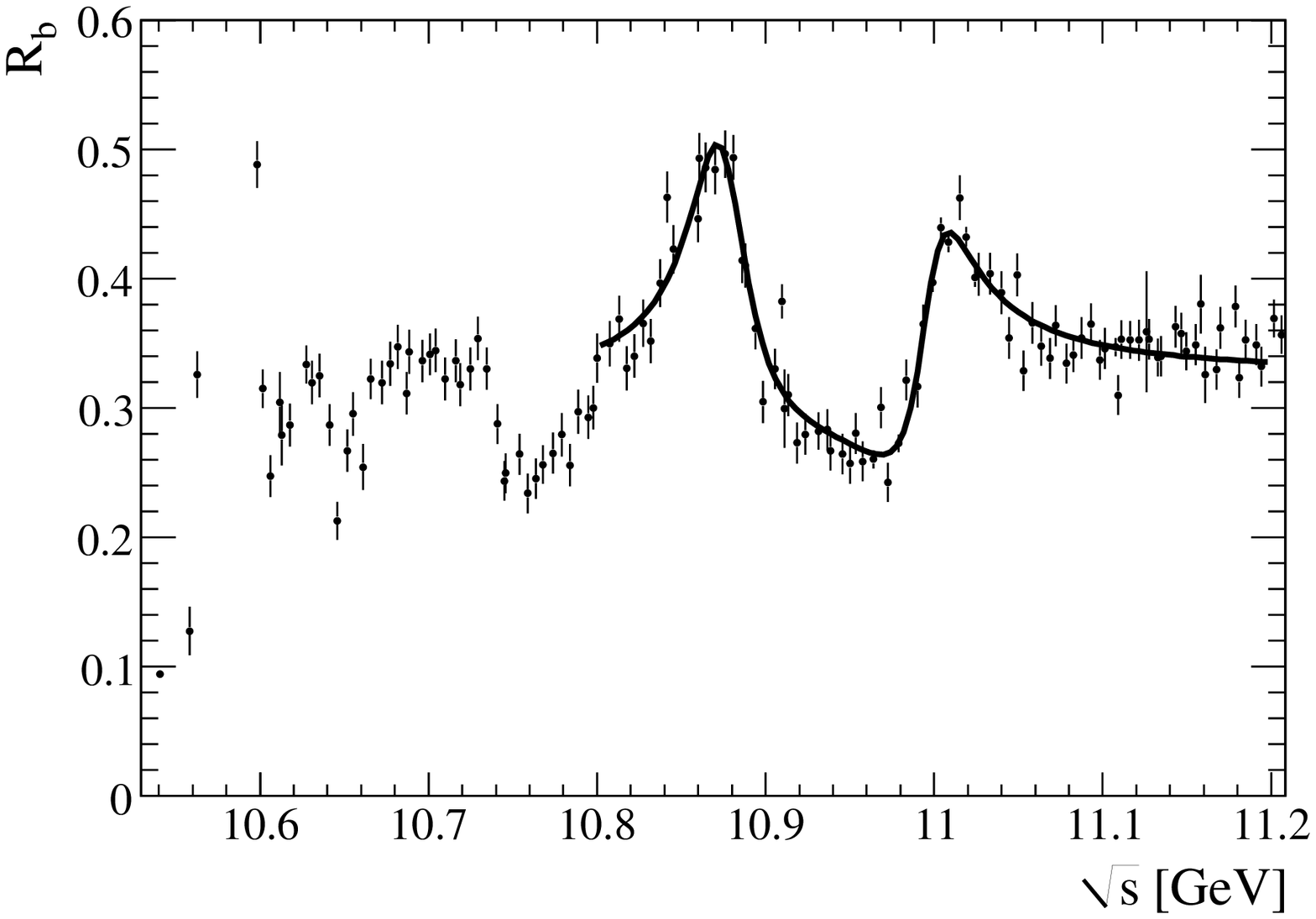}
\includegraphics[width=0.45\textwidth]{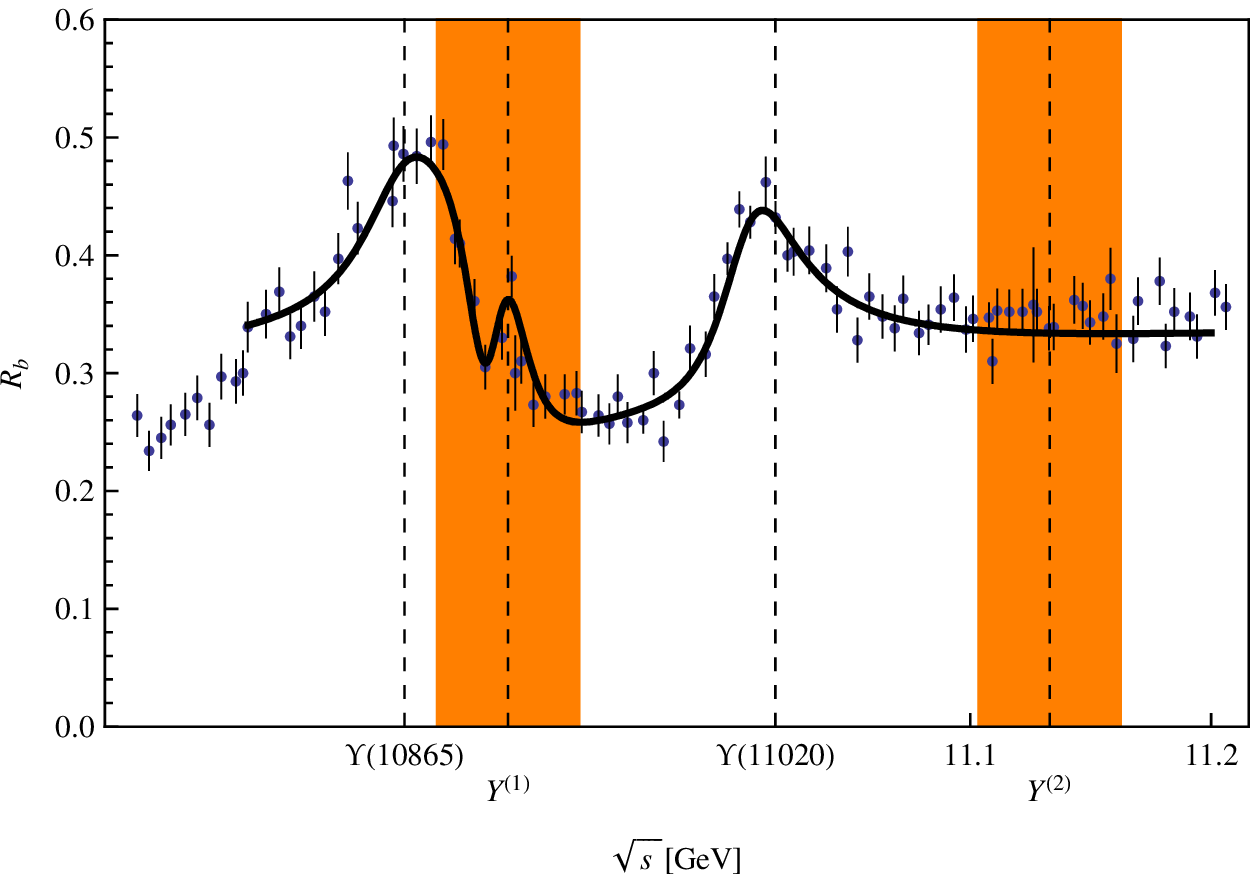}
\caption{\label{BABARPlots}
Measured $R_b$ as a function of $\sqrt{s}$ with the result of the fit with 2 Breit-Wigners described in
 \cite{:2008hx} (left frame). Reprinted from  Fig.~1 of B.~Aubert {\it et al.}  [BaBar Collaboration],
    Phys.\ Rev.\ Lett.\  {\bf 102}, 012001 (2009) [Copyright (2009) by the American
Physical Society]. The result of the fit with 4 Breit-Wigners
 described in the text is shown in the right-hand frame, where we have indicated the
location of the $\Upsilon (5S)$,  $\Upsilon (6S)$ and the tetraquark state
$Y_{[b,q]}$ (labelled as $Y^{(1)}$). The location of the next higher $J^{PC}=1^{--}$ state 
$Y_{[b,q]}^{(2)}$ (labelled as $Y^{(2)}$) is also shown. The shaded bands around
the mass of $Y^{(1)}$ and $Y^{(2)}$ reflect our theoretical uncertainty in the masses. }
\end{figure}

BaBar has recently reported the $e^+ e^- \to b \bar{b}$ cross section measured in a dedicated 
 energy scan  in the range $10.54$ GeV and $11.20$ GeV taken in steps of 5 MeV~\cite{:2008hx}.
Their measurements are shown in  Fig.~\ref{BABARPlots} (left frame) together with the result of the
 BaBar fit, the details of which are described in their paper and which were also made available to
us~\cite{Faccini:2009}. Their fit model of the $R_b$-data contains the following ingredients:
a flat component representing the $b\bar{b}$-continuum states not interfering with resonant decays,
called $A_{nr}$, added incoherently to a second flat component, called $A_r$, interfering with two relativistic
Breit-Wigner resonances, having the amplitudes  $A_{10860}$, $A_{11020}$ 
and strong phases, $\phi_{10860}$ and  $\phi_{11020}$, respectively. Thus,
\begin{equation}
\label{babarfit-simple}
\sigma (e^+e^-\to b \bar{b}) = \vert A_{nr}\vert^2 + \vert A_r + A_{10860}e^{i \phi_{10860} } BW(M_{10860}, \Gamma_{10860})
+ A_{11020}e^{i \phi_{11020} } BW(M_{11020}, \Gamma_{11020})\vert^2~,
\end{equation}
with $BW(M,\Gamma)=1/[(s-M^2) +i M \Gamma]$. The results summarised in their Table II for the masses
and widths of the $\Upsilon(5S)$ and $\Upsilon(6S)$ differ substantially from the corresponding
PDG values~\cite{Amsler:2008zzb}, in particular, for the widths, which are found to be
$43 \pm 4$ MeV for the $\Upsilon(10860)$, as against the PDG value of $110 \pm 13$ MeV,
and $37\pm 2$ MeV for the $\Upsilon(11020)$, as compared to $79\pm 16$ MeV in PDG.
 As the systematic errors  from the various thresholds are not taken into account,
this mismatch needs further study. 
The fit shown in Fig.~\ref{BABARPlots} (left frame) is not particularly impressive
having a $\chi^2/{\rm d.o.f.}$ of approximately 2. In particular, the data points around 10.89 GeV and
 11.2 GeV lie systematically above the fit. In our analysis of the BaBar data, we were able to reproduce
 these features,
but also found that the fit-quality can be improved somewhat at the expense of
 strong phases $\phi_{10860}$
and $\phi_{11020}$, which come out different than the ones reported by BaBar~\cite{:2008hx}.
 We do not show this
fit here as the resulting $R_b$-line-shape is close to the one shown in the BaBar publication and reproduced here.

We have repeated the fits of the BaBar $R_b$-data, modifying the fit model in
 Eq.~(\ref{babarfit-simple})
 by taking into account
two additional resonances, corresponding to the masses and widths of $Y_{[b,l]}$ and $Y_{[b,h]}$. Thus,
 formula \eqref{babarfit-simple} is extended by two more terms 
\begin{equation}
A_{Y_{[b,l]}}e^{i \phi_{Y_{[b,l]}} } BW(M_{Y_{[b,l]}}, \Gamma_{Y_{[b,l]}})
\en\en\en\textnormal{     and     } \en\en\en
A_{Y_{[b,h]}}e^{i \phi_{Y_{[b,h]}} } BW(M_{Y_{[b,h]}}, \Gamma_{Y_{[b,h]}}),
\end{equation}
which interfere with the resonant amplitude $A_{r}$
  and the two resonant amplitudes for $\Upsilon (5S)$ and $\Upsilon (6S)$
 shown in 
Eq.~(\ref{babarfit-simple}). We use the same non-resonant amplitude $A_{nr}$ and $A_r$ 
as in the BaBar analysis~\cite{:2008hx}.
The resulting fit is shown in Fig.~\ref{BABARPlots} (right frame).
Values of the best-fit parameters are shown in Table \ref{fitvalues}, from where one see that
 the masses of the $\Upsilon(5S)$ and $\Upsilon(6S)$ and their respective full widths 
 from our fit are almost identical to the values obtained by BaBar~\cite{:2008hx}. 
 However, quite strikingly,  a third resonances is seen in the
 $R_b$-line-shape at a mass of $\BBfitmassY$~GeV,
 tantalisingly close to the $Y_b(10890)$-mass in the Belle measurement
of the cross section for $e^+e^- \to Y_b(10890) \to \Upsilon(1S,2S)\; \pi^+\pi^-$, and a width 
of about $\BBfitwidthY$~MeV. 
In the region around 11.15 GeV, where the $Y^{(2)}_{[bq]}$ states are expected, our fits
 of the BaBar $R_b$-scan do not show a resonant structure due to the large decay widths
 of the states
$Y^{(2)}_{[bq]}$. The resulting $\chi^2/{\rm d.o.f.} = \chsqfit$ with the 3 Breit-Wigners shown 
in Fig.~\ref{BABARPlots} (right frame) is  better
than that of the BaBar fit.~\cite{:2008hx}.
 A Belle $R_b$-scan will greatly help to confirm or refute the existence of the state
$Y_{[bq]}$ visible in the analysis presented here. As the decays
 $Y_{[bq]} \to B_s^{(*)} \bar{B}_s^{(*)}$ are not allowed, restricting the final states in
 $R_b$ to the $ B_q^{(*)} \bar{B}_q^{(*)}$ ($q=u,d$), into which $Y_{[bq]}$ decay,
 will reduce the background to the $Y_{[bq]}$ signal.  
 It will be crucial to
check that the characteristics of $Y_{[bq]}$ (mass, full width and the electronic width)
 match those of the $Y_b$, measured in the exclusive 
process $e^+e^- \to Y_b \to \Upsilon(1S, 2S)\; \pi^+ \pi^-$. This may  solve one of the outstanding
 mysteries in the $\Upsilon(nS)$ physics.

 The quantity $\mathcal{R}_{ee}(Y_b)$ in \eqref{ratiodef} is given by the ratio of the two amplitudes
 $A_{Y_{[b,l]}}$ and $A_{Y_{[b,h]}}$, which also fixes the mixing angle $\theta$. From our fit,
 we get 
 \begin{equation}
 \mathcal{R}_{ee}(Y_b)=1.07 \pm 0.05,
 \end{equation}
yielding  
\begin{equation}
\theta=-19 \pm 1 ^{\circ}\en\en\en\textnormal{and}\en\en\en \Delta M=5.6 \pm 2.8\;\textnormal{MeV},
\end{equation}
for the  mixing angle and  the mass difference between the eigenstates, respectively.

The $R_b$-analysis in the tetraquark picture can be used to determine  $\kappa$. The
procedure how to do this requires 
some discussion. $\kappa$ can be determined from the theoretically estimated total decay widths of the
$Y_{[b,q]}$ states and the corresponding result from the $R_b$-fit. However, the estimated decay width of the
$Y_{[b,q]}$ is based on the input $\Gamma [\Upsilon(5S)]=110 \pm 13$ MeV from the PDG.
 The BaBar fit  and ours, on the other hand, yield a lot smaller value for this decay width
 (see, Table~\ref{fitvalues}). 
 To avoid the dependence on the absolute value of $\Gamma_{\rm tot}[\Upsilon (5S)]$,
it is safer to determine $\kappa$ from the ratios of the theoretical decay widths
$\Gamma_{\rm tot}(Y_{[b,q]})/\Gamma_{\rm tot}[\Upsilon(5S)]_{\rm theory}=  (88 \pm 16)\kappa^2/(110 \pm 13)$,
 and the corresponding ratio of these widths obtained from the fit of the $R_b$-data,
$\Gamma_{\rm tot}(Y_{[b,q]})/\Gamma_{\rm tot}[\Upsilon(5S)]_{\rm fit}= (28 \pm 2)/(46 \pm 8)$. This yields 
(adding the errors in quadrature):  
 \begin{equation}
 \kappa=\sqrt{\frac{110 \pm 13}{88\pm 16}
 \frac{28\pm 2}{46\pm 8}}=\kappaval \pm \kappavalError,
 \end{equation}
which is in the right ball park expected from the Lattice QCD estimates of the
 same~\cite{Alexandrou:2006cq}.
For the mass eigenstates $Y_{[b,l]}$ and  $Y_{[b,h]}$, the electronic widths $\Gamma_{ee}(Y_{[b,l]})$
and $\Gamma_{ee}(Y_{[b,h]})$ are given by 
$\Gamma_{ee}(\theta)=0.4\; \kappa^2 Q(\theta)^2$ keV, as already stated. With the above determination of
$\kappa$ and $\theta$, we get  
\begin{equation}
\Gamma_{ee}(Y_{[b,l]})=0.09\pm 0.03 \; \textnormal{keV}\en\en\en\textnormal{and}\en\en\en \Gamma_{ee}(Y_{[b,h]})=0.08\pm 0.03\; \textnormal{keV}.
\end{equation}
\vspace*{-5mm}
\begin{table}[h!]
\caption{Fit values of the masses, decay widths (both in MeV) and the strong phases $\phi$ (in radians).
    }
\label{fitvalues}
\begin{center}
\begin{tabular}{|l|l|l|l|}
\hline
                  & $M  [ MeV ] $                   & $\Gamma  [ MeV ]$       &  $\varphi$ [rad.]  \\ \hline
$\Upsilon(5S)$    & $ 10864\pm 5 $                  & $ 46 \pm 8      $       &   $1.3 \pm 0.3 $       \\ \hline
$\Upsilon(6S)$    & $ 11007\pm 0.3$                 & $ 40 \pm 2          $   &   $ 0.88\pm 0.06  $     \\ \hline
$Y_{[b,l]}$ & $ \BBfitmassYMeV-\Delta M/2 \pm 2 $& $ \BBfitwidthY \pm 2  $ &   $4.4 \pm 0.2  $      \\ \hline
 $Y_{[b,h]}$& $ \BBfitmassYMeV+\Delta M/2 \pm 2 $& $ \BBfitwidthY \pm 2  $ &  $ 1.9\pm 0.2$
\\ \hline\hline
\end{tabular}%
\\[0pt]
\end{center}
\end{table}
In conclusion, we have presented a case for the observation of a hidden  $b\bar{b}$ tetraquark states in
 the BaBar $R_b$-scan~\cite{:2008hx}.
 Our analysis is compatible with a $J^{PC}=1^{--}$ state $Y_{[bq]}(10900)$ having a width of
about 30 MeV. A scan of $R_b$ in finer energy steps should be able to resolve the structure
seen at this mass in terms of two mass eigenstates, split by about 6 MeV. The electronic widths are
estimated to be between 50 and 120 electron volts.  Other possible manifestations of tetraquarks
have been discussed in the literature~\cite{Hou:2006it,Karliner:2008rc} and
a dynamical model for the decays $Y_b (10890) \to \Upsilon(1S,2S) \pi^+\pi^-$
 is presented in~\cite{Ali:2009es}. 

{\bf Acknowledgments}

We thank the BaBar collaboration and the American Physical Society for their permission to
show Fig.~\ref{BABARPlots} (left frame) published in~\cite{:2008hx}. Helpful discussions
with Riccardo Faccini and Antonello Polosa are gratefully acknowledged. We also thank
Riccardo Faccini for providing us the fit program used in the BaBar analysis and we are
grateful to Alexander Parkhomenko for reading the manuscript and pointing out several typos
and notational inconsistencies in the first version of this paper. This work has been
partially supported by funds provided by the ENSF, Trieste, Italy. Two of us (I.A. and
M.J.A.)  would like to thank DESY for the hospitality during the summer 2009, where
this work was done. C.H. wants to  thank Benjamin Lutz for discussion on some technical
aspects of the fits.


\end{document}